\documentclass[preprint2]{aastex}
	\usepackage{natbib}
	\usepackage{subfigure}
	\usepackage{amsfonts}
	\usepackage{multirow}
	\usepackage{tabularx}
	\usepackage{diagbox}
	\usepackage{tcolorbox}
	\usepackage{color}
	\usepackage{graphicx}
	\usepackage{amsmath}
	\usepackage{amssymb}
	\usepackage{mathrsfs}
	\usepackage{makecell}
\newcommand{\m}{\ensuremath{\,{\rm m}}}
\newcommand{\km}{\ensuremath{\,{\rm km}}}

\newcommand{\MHz}{\ensuremath{\, {\rm MHz}}}

\newcommand{\mat}[1]{\boldsymbol{\mathbf{#1}}}

\begin{document}

\title{An imaging algorithm for a lunar orbit interferometer array}

\author{Qizhi Huang\altaffilmark{1,2,3},  Shijie Sun\altaffilmark{1}, Shifan Zuo\altaffilmark{1,2},
Fengquan Wu\altaffilmark{1}, Yidong Xu\altaffilmark{1},  Bin Yue\altaffilmark{1}, 
Reza Ansari\altaffilmark{3}, Xuelei Chen\altaffilmark{1,2,4}}
\altaffiltext{1}{Key Laboratory of Computational Astrophysics, National Astronomical Observatories, Chinese Academy of Sciences, Beijing 100101, China}
\altaffiltext{2}{School of Astronomy and Space Science, University of Chinese Academy of Sciences, 
Beijing 100049, China}
\altaffiltext{3}{Universit\'e Paris-Sud, LAL, UMR 8607, F-91898 Orsay Cedex, France $\&$ CNRS/IN2P3, F-91405 Orsay, France}
\altaffiltext{4}{Center for High Energy Physics, Peking University, Beijing 100871, China}
\email{E-mail: xuelei@cosmology.bao.ac.cn}	
%%%%%%%%%%%%%%%%%%%%%%%%%%%%%%%%%%%%%%%%%%%%%%%%%

\begin{abstract}
Radio astronomical observation below 30 MHz is hampered by the refraction and absorption of the ionosphere, and the 
radio frequency interference (RFI), so far high angular resolution sky intensity map is not available. 
An interferometer array on lunar orbit provides a perfect observatory in this frequency band: it is out of 
the ionosphere and the Moon helps to block the RFIs from the Earth.
The satellites can make observations on the far side of the Moon and then send back the data on the 
near side part of the orbit. However, for such array the traditional imaging algorithm is not applicable: the field of 
view is very wide (almost whole sky), and for baselines distributed on a plane, there is a mirror symmetry between
the two sides of the plane.  A further complication is that for each baseline, the Moon blocks part of the sky, but 
as the satellites orbit the Moon, both the direction of the baseline and the blocked sky change, so even imaging 
algorithms which can deal with non-coplanar baseline may not work in this case.
Here we present an imaging algorithm based on solving the linear mapping equations relating the sky intensity to 
the visibilities. We show that the mirror symmetry can be broken by the three dimensional baseline 
distribution generated naturally by the precession of the orbital plane of the satellites. The algorithm is applicable and good maps could be 
reconstructed, even though for each baseline the sky blocking by the Moon is different.
We also investigate how the map-making is affected by inhomogeneous baseline distributions. 
\end{abstract}


\maketitle
%%%%%%%%%%%%%%%%%%%%%%%%%%%%%%%%%%%%%%%%%%%%%%%%%%

\section{Introduction}

Radio astronomy began at low frequency with the observation by Karl Jansky at 20.5 MHz \citep{1933Natur.132...66J},
and Grote Reber made many low frequency observations down to hectometer 
waveband over the years \citep{1994JRASC..88..297R}. However, ground based low frequency radio 
observation suffers from strong ionosphere refraction below $ 30 \MHz$ and absorption below $ 10 \MHz$. 
Furthermore, human generated radio frequency interference (RFI) at low frequency is almost omnipresent on the 
Earth due to reflection from the ionosphere. Except for strong emissions from solar radio bursts and 
planetary radio activities, there were only very few observations at the frequency below 30 MHz  
and the sky radiation in this frequency range 
is poorly known \citep{1968AJ.....73..717B,1977MNRAS.179...21C,1978AuJPh..31..561C}.
In recent years, the interest on low frequency radio astronomy were renewed, especially for the observation of 
the redshifted 21cm line from the Epoch of Reionization(EoR), cosmic dawn and the dark ages \citep{2012RPPh...75h6901P}.
Some ground-based experiments, such as the low frequency array (LOFAR)  \citep{2013A&A...556A...2V}
and Long Wavelength Array (LWA) \citep{2011MmSAI..82..664C}
have bandwidth coverage below 30 MHz. 
Nevertheless, most observations of these new arrays are made at higher frequencies, 
the lower band is mainly used for observation of the Sun and planets.
It is therefore highly desirable to make the low frequency astronomical observations from the space.

In the 1970s, the IMP-6 \citep{1973ApJ...180..359B}, Radio Astronomy Explore (RAE)-1 \citep{1974AJ.....79..777A} and
RAE-2 \citep{1975A&A....40..365A}  satellites made some low frequency radio observations from space. The data collected by 
these satellites showed that the Earth have strong radio emissions at the relevant band, but the Moon can shield the 
spacecraft from this emission of the Earth, so the far side of the Moon provides an ideal environment for low frequency 
radio observation.  However, limited by the technology available then, the sky map constructed from the 
RAE missions are of poor angular resolution\citep{1978ApJ...221..114N}, and there are also substantial differences in the 
spectrum measured by these missions \citep{Keshet:2004dr}.

Since the 1980s, a number of dedicated low frequency radio space mission concepts have been proposed (see 
e.g. \citealt{1990LNP...362.....K,1997RaSc...32..251B,2000GMS...119....1K} for a review of the early studies),
though so far none have been realized. However, some low frequency observations were made by instruments on board 
spacecrafts such as the WIND \citep{1995SSRv...71..231B} and CASSINI \citep{2004SSRv..114..395G}.

A mission on the Earth orbit (e.g. the solar radio mission 
SunRISE \citealt{2017EGUGA..19.5580L}) is economical and relatively simple in terms of technology, but 
the RFI from the Earth would be a major concern for  imaging the sky or even probing the dark ages.
An array on the Sun-Earth L2 point, e.g. ALFA \citep{1998ASPC..144..393J},  FIRST \citep{2009arXiv0911.0991B}, 
SURO-LC \citep{2013EPSC....8..279B}, allows all-time monitoring of the whole 
sky, but it is also constantly exposed to the radio emission from the Earth, though reduced in 
magnitude by the distance. Launching the satellites and maintaining the array 
configuration around the unstable L2 point, determining their relative positions, and transmitting the data back all 
require a lot of research and development and substantial amount of resources both on board the satellite and on ground. 
Radio observation from the far side lunar surface is another option (see reviews in 
\citealt{2000GMS...119..351K,2009NewAR..53....1J,2012ExA....33..529M,2012P&SS...74..156Z}). 
A  first experiment is to be carried out in 2018 by the Chang'e-4 (CE-4) lander \citep{2016AcAau.127..678W}.
For an array on the lunar surface, the imaging methods and tools developed for the ground-based radio 
astronomy could be 
readily  used. However, a relay satellite is required to transmit the data back to the Earth, and furthermore, 
supplying energy for the lander during the half-month long lunar night requires special power 
source such as radioisotope thermoelectric generator.

From technological point of view, making radio observations from the orbit around the Moon 
has a number of advantages compared with the other space options, which makes it easier to realize in 
near terms. The part of orbit behind the Moon provides perfect environment against the radio emissions from the Earth.
Unlike the lunar surface,  the orbital period is about two hours for an orbit of 300 km altitude,
so the solar power could be used. The data can be transmitted back to the Earth when it is on the near side, without the
need of relay satellite, and the complicated landing and deployment is also avoided. A single lunar satellite such as the 
proposed DARE experiment \citep{2017arXiv170200286P} could measure the global average spectrum, 
and also observe strong radio sources or even 
map the sky with low angular  resolution by using the Moon as a moving screen. A number of satellites may also form an 
interferometer array \citep{1992ecos.proc.1913B,Chen2005a} which could provide higher angular resolution. Several 
conceptual studies have been conducted, including the OLFAR \citep{2010cosp...38.2378B}, 
DARIS\citep{2010cosp...38.2364B} and 
the Discovering the Sky at Longest wavelength (DSL, \citealt{7500678}) missions. 

\begin{figure}[tbp]
\centering
\includegraphics[width=0.4\textwidth]{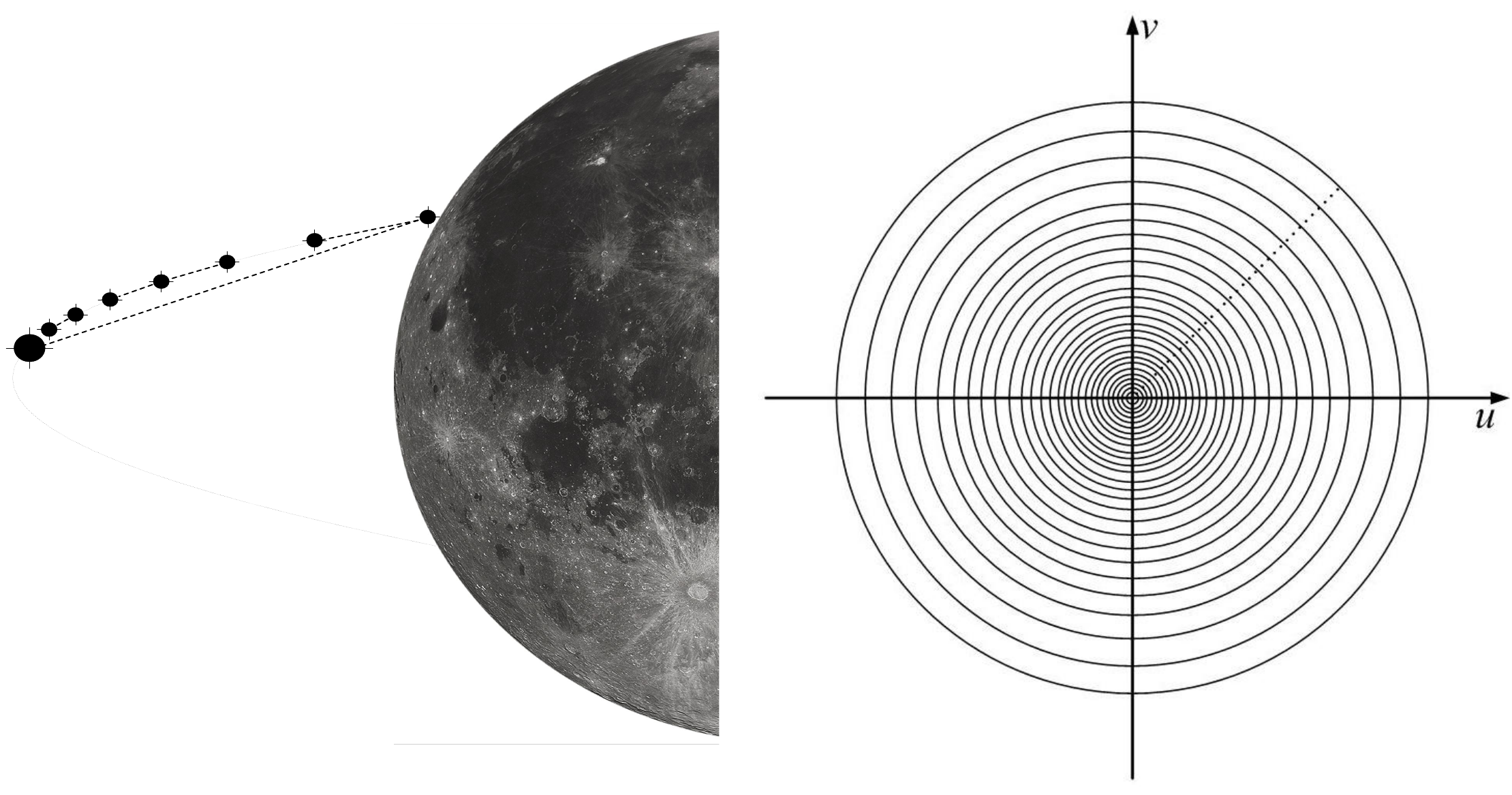}
\caption{Left: A linear array of formation flying satellites around the Moon, with some of the baselines connected. 
Right: The baselines between satellite pairs in the array swipe a number of concentric rings on 
the $(u,v)$ plane during one orbit.}
\label{fig:sketch}
\end{figure}

In the DSL concept a mother satellite with a number of daughter satellites
form a linear array on the same circular orbit of 300 km altitude and fly in formation around the Moon, 
and as they orbit around, the baseline vectors 
between the different satellites  also circle around (Fig.~\ref{fig:sketch}) . The mother satellite measures 
the angular positions of and distances to each of the daughter satellites constantly by optical/microwave devices, 
so that the array configuration can be determined at any moment. 
The daughter satellites will use the electrically short antenna on board to make interferometric observation of the sky at the 
long wavelengths when the array is on the far side of the Moon, the data from each daughter satellite is then 
sent to the mother satellite for real-time cross correlation, the generated visibilities will be stored and sent back
to the Earth when they are on the near-side of the orbit.  After many orbits, the visibility measurements are taken for many 
different baselines, and due to the precession of the orbital plane, these baselines have
a three dimensional space distribution. If during the observation, the sources have a constant brightness, 
by synthesizing these measurements,  the image of the sky can be made, in analogy to the rotational 
synthesis imaging on the ground.
A first technology demonstration experiment of the orbiter interferometry will be carried out by two satellites 
piggy-backing on the the CE-4 relay satellite rocket to be launched in 2018\citep{Zhang2017a}. 

So far the space mission studies have focused mainly on the general concept and system configuration.
As the orbital interferometer array measurements do include all information about the sky brightness, 
it is assumed that good images of the sky could be reconstructed from the visibilities taken on the orbit, though 
so far the imaging algorithm have not been investigated in detail. However, unlike the lunar surface array, which can use 
existing tools and methods developed for the ground-based radio astronomy, the image reconstruction from the 
wide field interferometer array moving on the three dimensional orbit is far from trivial. In the present paper, we study the 
imaging algorithm for a lunar orbit array mission such as the DSL.

 The rest of this paper is structured as follows. In the next section, we will present the general formalism for 
 the lunar array image synthesis. In Sec. 3, we make a simulation of the imaging process, and assess the various effects
 on image synthesis. Finally, we conclude in Sec. 4.   

\section{Imaging Algorithm}
 In this work, we limit our discussion to unpolarized sky brightness. 
 The visibility, which is the correlation between the signal from two array element  is given by
\begin{equation}
	V_{ij} = \int A_{ij} (\hat{k}) T(\hat{k})~ e^{-i \vec{k} \cdot \vec{r}_{ij}}  d^2 \hat{k},
	\label{eq:Vij}
\end{equation}
where $A_{ij}(\hat{k})$ is the combined  antenna beam pattern,  $T(\hat{k})$ is the sky temperature in the direction
$\hat{k}$, $\vec{k}=\frac{\omega}{c} \hat{k}$ is the wave vector, 
and $\vec{r}_{ij}=\vec{r}_i-\vec{r}_j$ is the baseline vector between the $i$-th and $j$-th elements respectively. 
The visibility can also be expressed in $\vec{u} \equiv (u,v,w)$ which are the baseline vector in units of the 
wavelength, conventionally $w$ denotes the component along the axis pointing to the phase reference point on the 
celestial sphere,  and $\hat{k}=(l,m,n)$, where  $(l,m,n)$ are the direction cosines with respect to the 
coordinate axes and $n=\sqrt{1-l^2-m^2}$, then
\begin{eqnarray}
	V(u,v,w) &=& \int \frac{{\rm d}l\, {\rm d}m}{n} A_{ij}(l,m)  T(l,m)   \nonumber\\
	&&\times~ e^{-i2\pi\left[ ul + vm + w(n-1) \right]}
\label{eq:Vuvw}
\end{eqnarray}

Algorithms of reconstructing sky image from the visibilities have been developed since the 
invention of radio interferometry, the simplest is the 2D Fourier transform method.
If the field of view is narrow ($l$$\sim$0, $m$$\sim$0 then $n$$\sim$1) and the antenna array is 
coplanar ($w$$\sim$0), Eq.~(\ref{eq:Vuvw}) is reduced to a 2D Fourier transformation
\begin{eqnarray}
	V(u,v) = \int {\rm d}l\,{\rm d}m ~A_{ij}(l,m)\, T(l,m)\, e^{-i2\pi(ul + vm)}
\label{eq:Vuv}
\end{eqnarray}
The sky intensity could be recovered from the visibilities by a two dimensional (2D) inverse Fourier transform.
But if the field of view is large, the $w$-term in Eq.~(\ref{eq:Vuvw}) can not be neglected.
A number of wide-field imaging formalisms have been 
developed, such as faceting \citep{1992A&A...261..353C}, the 3D Fourier transform \citep{1989ASPC....6..117S, 
1992A&A...261..353C, 1999ASPC..180..383P}, the W-Projection \citep{2005ASPC..347...86C, 2008ISTSP...2..647C}, 
A-Projection \citep{2013A&A...553A.105T} and W-Stacking \citep{2014MNRAS.444..606O}. 
To improve the quality of the image, iterative deconvolution algorithms such 
as ``CLEAN'' \citep{1974A&AS...15..417H} are applied to 
the dirty map\citep{2007astro.ph..1171S}. To better deal with the direction-dependent effect in 
calibration and imaging for wide field-of-view antenna encountered in low frequencies, refined methods such as
the ``software holography'' \citep{2009MNRAS.400.1814M} have been developed. However, most of these 
algorithms are developed for ground based arrays, where the baselines are still mostly distributed near a plane.
As we shall see below, this is not appropriate for the orbital array.

\subsection{The Mirror Symmetry}

\begin{figure}[htbp]
\centering
\includegraphics[width=0.4\textwidth]{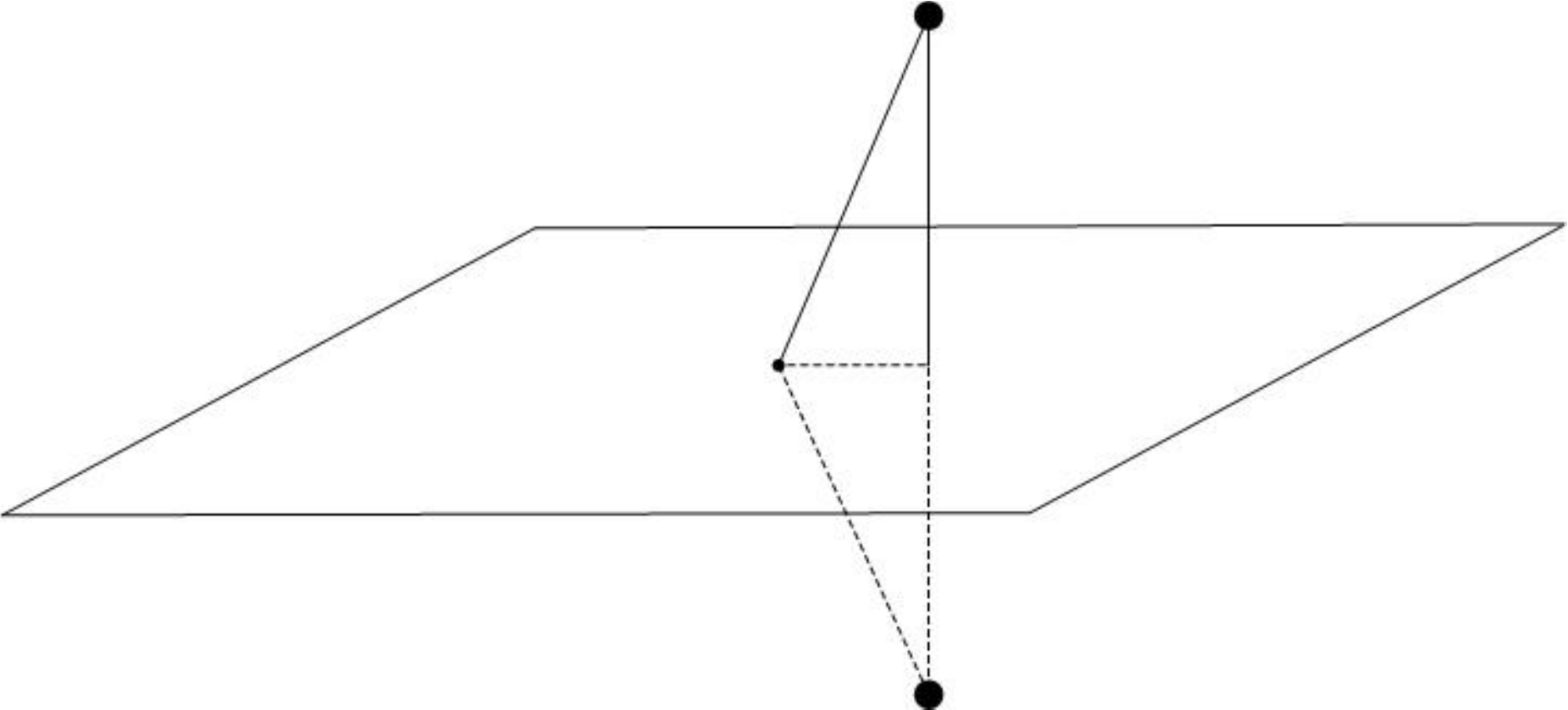}
\caption{Mirror symmetry with respect to the orbital plane. }
\label{fig:mirror}
\end{figure}

During one orbital period, the baselines may be regarded approximately as distributed on a single plane--the orbital plane.
 However,  unlike ground-based arrays, where the ground (Earth) serve to block half of the whole sky, in the lunar orbit array 
case, for a reasonably stable orbit, the Moon only blocks less than half of the sky at any time, and in fact it does not 
block the directions near the normal of the orbital plane. So there is a mirror symmetry with respect to the orbit plane:
for each direction on the celestial sphere, there is always a mirror symmetric point with respect to the 
plane (Fig.~\ref{fig:mirror}). For any baseline
on the plane, the pair of mirror-symmetric points always make the same angle, so the phase delay along the baseline 
would be the same for the two directions.The antenna will receive signal from both sides of the plane, and 
because of this mirror symmetry, we could not distinguish from which of the two sides a signal comes from.  

This degeneracy may be broken by employing antenna with asymmetric beam pattern. Indeed, in the usual 
application,  for the commonly used reflectors the backside response is several tens decibels below that of 
the main lobe, this mirror symmetry can almost be neglected.  
However, this require an antenna with size at least comparable with the wavelength.
For the long wavelength we are considering, which ranges from 10 meter (30MHz) to 
1 kilometer (0.3 MHz), this is impractical and the antenna is most likely some type of electrically short dipole 
(The RAE-2 satellite was equipped with several long antennas, the longest is the 229 m V-antenna,  
but it would be hard to equip all the micro- or nano- satellites making up the array 
elements with such long antennas). For such short dipole antenna the antenna beam pattern is 
very wide and point symmetric, only when the wavelength comes into the range of the antenna size
will the pattern becomes asymmetric, so it is not easy to use this method to distinguish the two sides of the plane. 
Another possible way to break the symmetry is to move the beam pattern in space by 
either mechanical rotation or electronic steering during the observation
but that would complicate the satellite design as well as the data processing. 

A more simple and elegant solution to the mirror symmetry-breaking 
problem in this case is to employ the three dimensional  (3D) distribution of baselines. This can be 
achieved by allowing the orbital plane to precess, which would occur naturally for most orbits. 
In the DSL case, for example, the height of the circular orbit is $300~\km$, the orbit inclination angle is $30^\circ$,  
and the precession period is 1.29 year \citep{7500678}, so over the time of a few months,
visibility data could be acquired over a 3D distribution of baselines. By synthesizing these measurements, the 
mirror symmetry is automatically broken. 
However, this do require a fully three-dimensional algorithm of image synthesis.

A number of algorithms are available for image synthesis with non-coplanar baselines. 
For example, from Eq.~(\ref{eq:Vuvw}), the sky intensity can in principle be obtained by a direct 3D inverse 
Fourier transform \citep{1989ASPC....6..117S, 1992A&A...261..353C, 1999ASPC..180..383P}.
\begin{eqnarray}
  &&A(l,m) T(l,m) \frac{\delta (n-n')}{n} = ~\nonumber\\ 
	&&  \int {\rm d}u\, {\rm d}v\, {\rm d}w ~ V(u,v,w) e^{i2\pi\left[ul + vm + wn')\right]} 
\end{eqnarray}
where  $\delta(n-n')$ is the Dirac $\delta$-function. 
However, this method or other previously developed ones are actually inapplicable in our case,
because as the array circles the Moon, the part of sky being blocked by the Moon also changes.
If we treat the Moon as a opaque screen without reflection, and neglect the small differences in the 
position of the satellites, we have 
\begin{eqnarray}
 V(\vec{b}) = \int A(\hat{k}) S(\hat{k}, \vec{R}) T(\hat{k})\, e^{-i \vec{k} \cdot \vec{b}}\, {\rm d}^2\hat{k}
\end{eqnarray}
where the screen function $S(\hat{k},\vec{R})=0$ for the part of sky $\hat{k}$ being blocked by the Moon 
and 1 for the part not blocked. $S$ depends on the satellite position $\vec{R}$ relative to the center of Moon. 
But for the linear array orbiting the Moon on circular orbit, however, the baseline vectors are almost the tangent 
vectors of the orbit, $\vec{b} \propto \vec{R} \times \vec{N}$, where $\vec{N}$ is the normal of the orbital plane, so 
$S$ depends on $\vec{b}$. This makes the direct 3D Fourier transform method invalid. The imaging algorithm must 
 be able to deal with  the changing blocking of sky by the Moon. 

Despite these complexities, the visibility data are linearly related to the sky brightness distribution, so the 
interferometric imaging problem can always be solved as a general linear inversion problem. 
This is also the approach we adopted in this paper.

\subsection{Brute-Force Map Making}
\label{sec:pixel}

If the array response is linearly related to the sky intensity (even though it varies all the time 
as in our case), the linear mapping relation could be inverted by brute-force to recover the sky image.
This method was used in the ground-based MITEoR experiment
\citep{2017MNRAS.465.2901Z}. Here we use this approach to make image from the data of the lunar orbit array.

Discretizing the integral over sky angles into 
a sum over sky pixels, 
\begin{eqnarray}
	V_{ij}(t) = \sum_\alpha^{N_{\rm pix}} B(\alpha,t)\, T(\alpha) \Delta\Omega
\label{eq:Vijn}
\end{eqnarray}
where $T(\alpha)$ is the discrete sky map, $B(\alpha,t)=A_{ij}(\alpha,t)\, S(\alpha,t) e^{-i \vec{k_\alpha} \cdot \vec{r}_{ij}(t)}$ 
is the discrete complex response, $A_{ij}(\alpha,t)$ is the beam and $S(\alpha,t)$ the screening function of the Moon, 
$\Delta\Omega$ is the pixel angular size. Written in matrix form, and also including noise, 
\begin{eqnarray}
	\mat{V} = \mat{B}\, \mat{T} +\mat{n}.
\label{eq:matrixV}
\end{eqnarray}
$\mat{B}$ is a $(N_{\rm bl}\cdot N_{\rm t}) \times N_{\rm pix}$ matrix, where $N_{\rm bl}$, $N_{\rm t}$ and $N_{\rm pix}$ 
are the number of baselines, observation time points and the  number of pixels respectively. The
vector $\mat{T}$ is the discrete sky map with dimension $N_{\rm pix}$,  the visibility 
 $\mat{V}$ has the dimension $(N_{\rm bl}\cdot N_{\rm t})$, $\mat{n}$ is the random noise, with the noise covariance matrix
 $\langle \mat{n} \mat{n}^\dagger \rangle = \mat{N}$. 
 
 In this work we shall pixelize the sky with 
 HEALPix \citep{2005ApJ...622..759G},  $N_{\rm col}=N_{\rm pix}=12 n_{\rm side}^2$ 
is determined by $n_{\rm side} \equiv 2^p$ where $p$ is a non-negative integer. 
For a pixelized sky map with $1^{\circ}$ pixel size, $n_{\rm side}=64$.

The minimum variance estimator of $\mat{T}$ is 
\begin{equation}
\hat{\mat{T}}=(\mat{B}^\dagger \mat{N}^{-1} \mat{B})^{-1} \mat{B}^\dagger \mat{N}^{-1} \mat{V} 
\equiv \mat{B}^{-1} \mat{V}.
\end{equation}
For simplicity, here we assume the noise is uniform, i.e. $\mat{N} \propto$ the identiy matrix $\mat{I}$. 
We compute $\mat{B}^{-1}$ by the Moore-Penrose pseudo-inverse method, i.e. making a 
Singular Value Decomposition (SVD) of the matrix $\mat{B}$,
\begin{equation}
\mat{B} = \mat{U} \mat{\Sigma} \mat{W}^\dagger
\end{equation}
where $\mat{\Sigma}$ is a diagonal matrix. Picking out the substantially non-zero singular values 
$\mat{\bar{\Sigma}} = \mat{\Pi} \Sigma$ where $\mat{\Pi}$ is projecting matrix, the pseudo-inverse is 
\begin{equation}
	\mat{\bar{B}}^{-1} = \left( \mat{U} \mat{\bar{\Sigma}}^{-1}\mat{W}^{\dagger} \right)^{\dagger}, 
\label{eq:svd}
\end{equation}
where $\bar{\mat{\Sigma}}^{-1}$ is obtained by taking the 
inverse of each non-zero element in the matrix diagonal, while setting the other elements zero. 
The very small singular values may induce both numerical errors as well as  large noise contribution during inversion. 
In order to maintain stability of computation, one may set absolute and relative thresholds as a fraction of 
the largest singular value \citep{Zhang:2016whm,Zhang:2016miz}. The threshold value is usually set empirically.
In this work we make automatic adjustment by first sorting the singular values in a decreasing order,
then choose the threshold by requiring the cumulative ratio of singular values
$\sum_{i=1}^{N_{\rm thr}} \lambda_i \big/ \sum_{\rm all}^N \lambda_i =R_{\rm th}$, where we take the threshold
ratio $R_{\rm th}=0.99$.  Roughly speaking, the first $N_{\rm thr}$ singular values components we retain accounts 
for 99\% of total information \citep{deOliveiraCosta:2008pb}.  Our test show that this recipe works well and leads to stable
computation. 

We can use the point spread function (PSF) 
to assess the quality of the reconstruction.  The map we made is related to the original map by
\begin{equation}
\hat{\mat{T}}= \mat{P} \mat{T}
\end{equation}
The  point spread matrix $\mat{P}$ is given by
\begin{eqnarray}
	\mat{P} \equiv \mat{\bar{B}}^{-1} \mat{B} = \left( \mat{U} \mat{\bar{\Sigma}}^{-1}\mat{W}^{\dagger} \right)^{\dagger} 
		\textbf{U}\, \mat{\Sigma}\, \textbf{W}^{\dagger}
\label{eq:psf}
\end{eqnarray}
Obtaining the Moore-Penrose pseudo-inverse consumes the majority of computation for the
linear system solution.

\subsection{Spherical Harmonic Expansion}
\label{sec:ourmethod}
We may also decompose the sky intensity and beam function in spherical harmonics,  
which in some cases could reduce the amount of computation.
The spherical harmonic expansions of the complex beam $\mat{B}$ and the sky intensity $\mat{I}$ are
\begin{eqnarray}
	B(\theta,\varphi,t) &=& \sum_{l=0}^{\infty} \sum_{m=-l}^l \mathcal{B}_{lm}(t) Y_{lm}(\theta,\varphi) \\
	T(\theta,\varphi)   &=& \sum_{l=0}^{\infty} \sum_{m=-l}^l \mathcal{T}_{lm} Y_{lm}(\theta,\varphi) 
\label{eq:Ilm}
\end{eqnarray}
As the sky brightness is real, $\mathcal{T}_{lm} = (-1)^m \mathcal{T}_{l,-m}^*$.
In practice, we may sum over $l$ up to a maximum mode $l_{\rm max}$.
With the summation theorem of the spherical harmonics, 
Eq.~(\ref{eq:Vij}) can be rewritten as 
\begin{eqnarray}
	V_{ij}(t)   &=& \sum_{l=0}^{l_{\rm max}} \sum_{m=-l}^l  (-1)^m \mathcal{B}_{l,-m}(t) \mathcal{T}_{lm} , \\
	V_{ij}^*(t) &=& \sum_{l=0}^{l_{\rm max}} \sum_{m=-l}^l  \mathcal{B}_{l,m}^*(t) \mathcal{T}_{lm}.
\end{eqnarray}
and in a matrix form
\begin{eqnarray}
	\mat{V} = \mat{\mathcal{B}\, \mathcal{T}}
\label{eq:matrixV2}
\end{eqnarray}
For an antenna array with angular resolution $\theta_0$, we require in the reconstruction
$l_{\rm max} \geq \pi/\theta_0$.
For a $1^{\circ}$ map, we require $l_{\rm max}=180$. 
With very wide beams, as the one discussed here for the short dipoles, the beam patterns 
in the spherical harmonics space $(l,m)$ plane are localized with a small extension. The overall $B$ matrix
would be sparse and can be inverted more easily.
%%%%%%%%%%%%%%%%%%%%%%%%%%%%%%%%%%%%%%%%%%%%%%%%%%

\section{Simulations}

We apply here the map reconstruction method described above to the case of simulated
data from a lunar interferometer array. We shall consider a simple case, where the antenna beam is 
assumed to be omni-directional and uniform. Generalization 
to the case of short dipoles is straightforward. 

\subsection{Input Map}

\begin{figure}[htbp]
\centering
\includegraphics[width=0.4\textwidth]{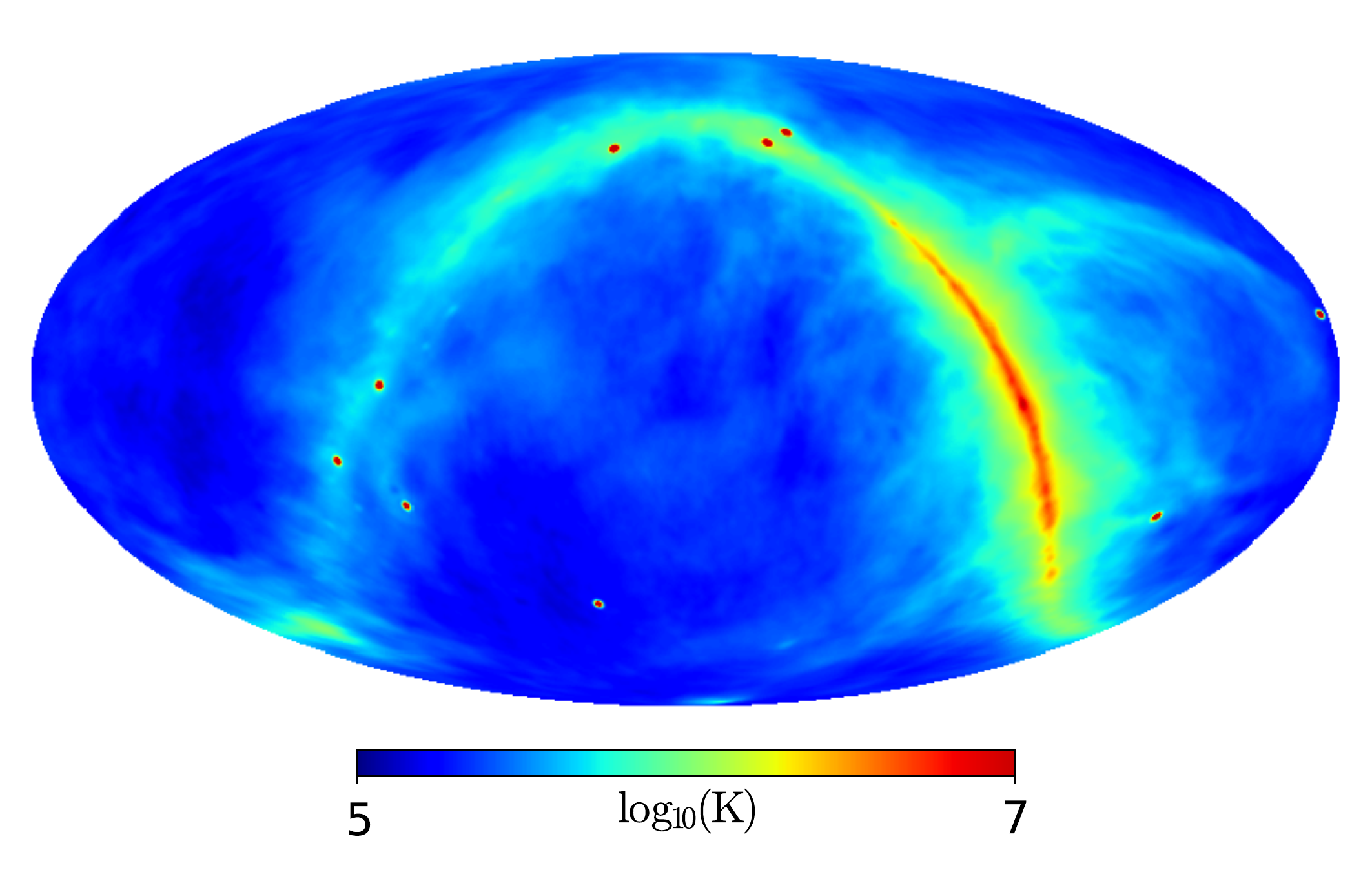}
\caption{The diffuse component map for simulation input, obtained by extrapolating the GSM model to 10 MHz, 
with $1^{\circ}$ resolution, shown in logarithmic scale.
}
\label{fig:inmap-diffuse}
\end{figure}

The input map we use is based on the improved Global Sky Model 
(improved GSM, \citealt{deOliveiraCosta:2008pb,2017MNRAS.464.3486Z}), 
which is derived from 29 sky maps at different frequencies 
by applying the iterative Principal Component Analysis (iterative PCA) algorithm and then extrapolated 
at the required frequency. In this paper we choose 10 MHz for all our computations. 
In the GSM map, the bright point sources have all been subtracted out. Here just to show on the same 
map how the point and diffuse sources behave under reconstruction, 
we added nine of the brightest radio sources, which are Cassiopeia A, Cygnus A, Cygnus X, 
Taurus A (Crab nebula), Orion A, Fornax A, Centaurus A, the Rosette nebula and Virgo A, 
and  smoothed them to $1^{\circ}$ to be consistent with the improved GSM map. We do not claim this map 
is a faithful representation of the sky, because many other point sources are not included.
Rather, it is used to show how the reconstruction program works.
We pixelize the map using HEALPix \citep{2005ApJ...622..759G}, 
with $n_{\text{side}}=512$, corresponding to a pixel size of $0.115^{\circ}$. 
The map is shown in Fig.~\ref{fig:inmap-diffuse}, the galactic plane is the most prominent feature in the map.

\subsection{The orbits and baselines}
\label{sec:procession}
Here we shall consider a simple case of uniformly sampling the 
visibilities for a static sky. In reality, the sky may not be static, in particular the Sun and planets such as the Jupiter could have
significant activity during the mission. However, their effect may in principle 
be reduced by using only the data taken when the Sun and Jupiter are non-active or be shielded by the  Moon. Alternatively,
more sophisticated algorithm may be developed to extract their contribution from the visibilities. For simplicity here we
just ignore the variable sources. The visibilities obtained by the orbiting array may also be non-uniformly distributed, 
as the observation would be made in the shielded zone, and the operation would
also be affected in practice by the available power, data storage and downlink resources, but these effects varies according 
to the specific condition of the mission, while in this study we are just interested in the general features of the 
imaging making process, so we shall not consider these complications.

The Moon's equator and its path of revolution around the Earth are all very close to the ecliptic plane, with only a few degree's
difference. As discussed earlier, the orbital plane precession is required in order 
to break the mirror symmetry with respect to a single
plane. A 3D baseline coverage could be achieved over a few months. In this simulation, we 
assume the inclination angle of the orbital plane with respect to the lunar equator (approximately the 
same as the ecliptic plane) is  $30^{\circ}$, the precession rate is relatively fast for this inclination angle. 

\begin{figure}[htbp]
\centering
\includegraphics[width=0.4\textwidth]{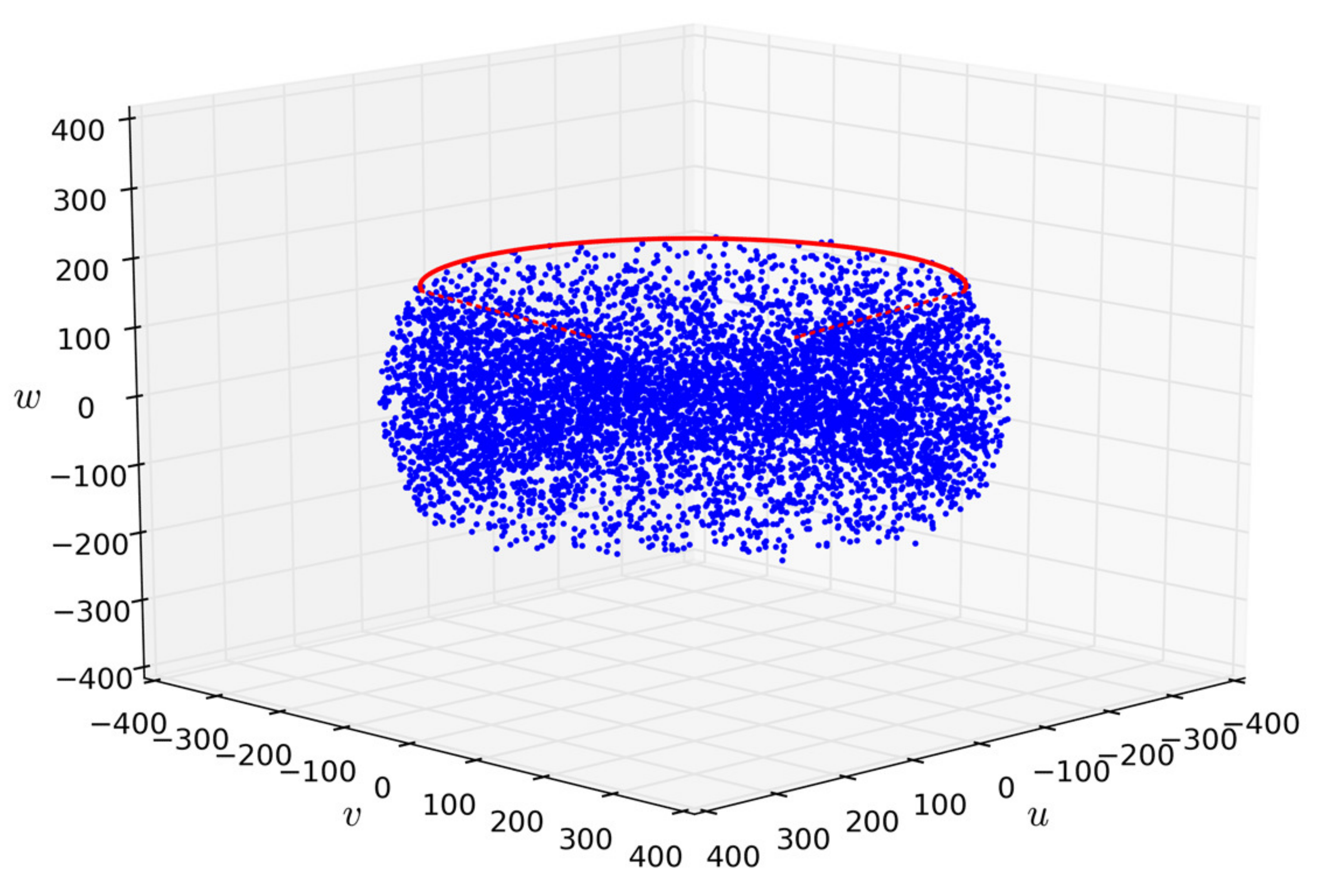}
\includegraphics[width=0.4\textwidth]{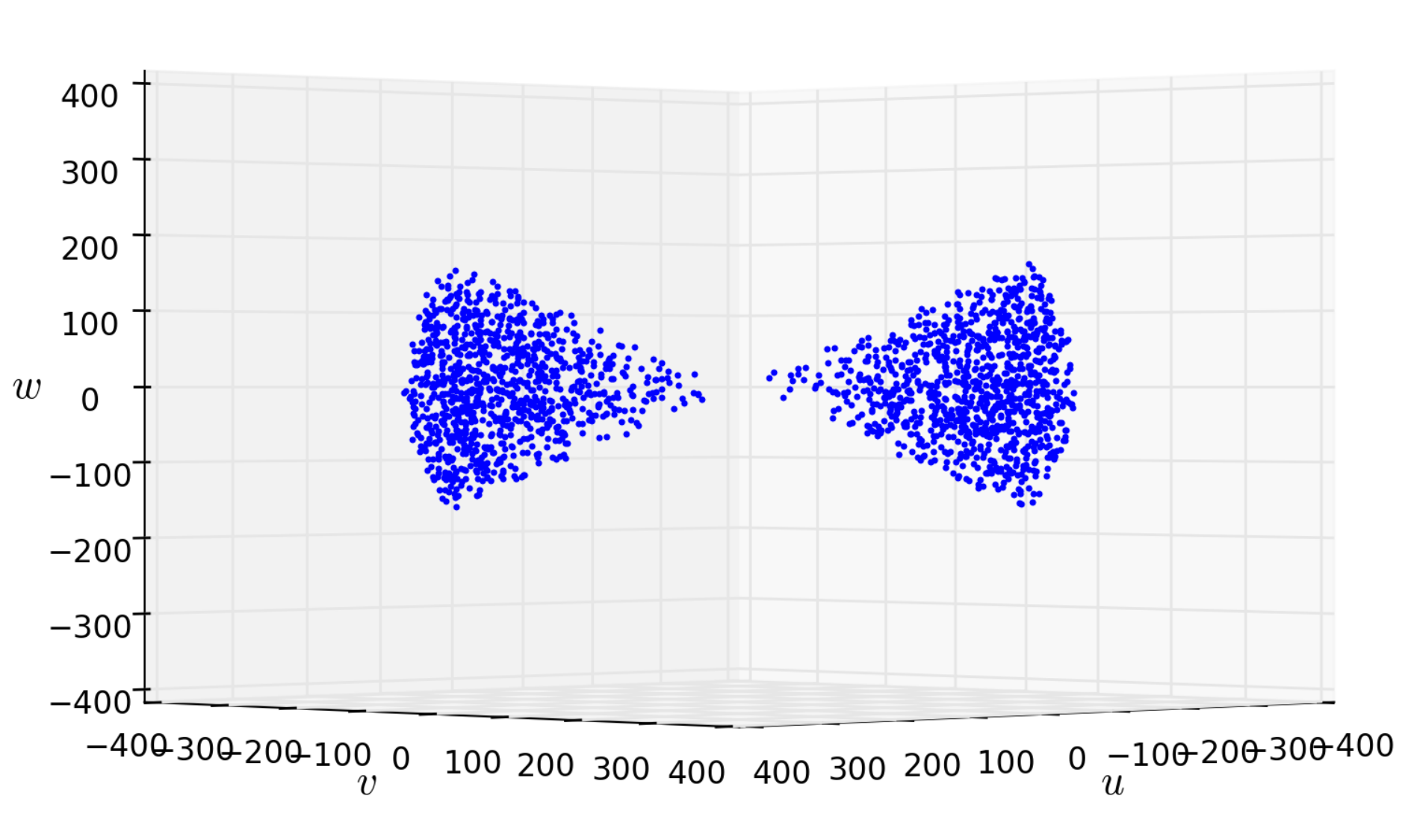}
\caption{ Top: the 3D distribution of baselines (blue points) as viewed from an angle of $15^\circ$ above the equator, 
we also added a few red lines and curve to help delineate the outline of the distribution; 
Bottom: a cros ssection through the center. }
\label{fig:blcover}
\end{figure}

We assume the array is made of a number of satellites orbiting on the same circular orbit, with different inter-satellite
distances, so that on each orbit plane the $(u,v)$ coverage is a series of concentric rings, as shown in Fig.~\ref{fig:sketch}. 
If the distances between the satellites vary a little bit, the rings will get some thickness. 
In this simulation the baselines are uniformly generated, but
in real mission, the satellites might be spaced logarithmically, then the 
density would not be uniform, but we ignore such complications in this work.
With the precession of the tilted orbital plane of the satellites, a three dimensional figure is produced, 
which may be described as a sphere with cones on the top and bottom removed, the angle of the cone equals the 
inclination angle of the orbit, i.e. $30^\circ$. We  show the a mock sample in the top panel of Fig.~\ref{fig:blcover}, and also
a vertical cross section through the center of this figure in the bottom panel.

In the DSL \citep{7500678}, the inter-satellite distance ranges from 0.1 km to 100 km. 
Longer baselines are technically more challenging as it requires higher precision of angular and 
distance measurement at longer distances, and also the sampling rate of the 
correlator must be increased, because the longer baselines will swipe through the $(u,v,w)$ space with 
faster speeds, hence the fringes move at higher rates.
On the other hand, the angular resolution of the observation is intrinsically limited by the 
scattering of interstellar medium (ISM) and interplanetary medium (IPM) \citep{2009NewAR..53....1J}. 
The $100 \km$ baseline yields $\sim$1 arcminute resolution at 10 MHz, which is about the limit allowed
by the ISM and IPM scattering. 

In the present study, we shall consider the baselines of 1--10 km. We choose this smaller range because in the
present study we are primarily interested in demonstrate that the brute-force method could make 
good maps of the sky with 3D distributed baselines. Also, the sky map with degree angular resolution is readily 
available for our simulation \citep{deOliveiraCosta:2008pb,2017MNRAS.464.3486Z}, but currently there is no good map with 
finer resolutions at low frequencies. This is also the distance for the CE-4 lunar orbit 
interferometer pilot experiment \citep{Zhang2017a}. 
For concreteness, we shall consider the observation at 10 MHz ($\lambda =30 \m$), so $33.3 < |\vec{u}| < 333$.
About $8 \times 10^3$ $(u,v,w)$ points are generated with uniform random distribution. This number of points are sufficient
to make good reconstructed map at our angular resolution, while the computation time required is also not too long.

The Moon blocks part of the sky. We treat the Moon as a perfect opaque sphere without reflection, 
the deviation from sphere and 
the rugged terrain of the real Moon are ignored in this paper. 
For a lunar satellite with low altitude orbit, the screened region has an angular size of 
\begin{eqnarray}
	\theta_m \approx 2 \arcsin\left( \frac{R_{\rm m}}{R_{\rm m}+h} \right)
\label{eq:shade_size}
\end{eqnarray}
where $R_{\rm m}=1737.1 \km$ is the average radius of the Moon, $h$ is the height of the satellite to the lunar surface. 
For $h=300 \km$ as in the case of DSL, $\theta_m = 117^\circ$, and the Moon extends an angular area of 3.0 steradian, 
about a quarter of the whole sky.

As the satellites circles the Moon, the direction of the screened region varies with time. 
If the satellite's orbit is elliptical, the angular size of the screened region will also vary with time, but as we assumed a 
circular orbit it is fixed. In this first treatment we ignored effects of reflection, thermal emission and 
diffraction of radio waves by the Moon.  We also ignored the slight different blocking of the sky 
for different satellites due to the difference in their spatial position. 
Such effects will be investigated in subsequent studies.  The visibilities are then generated using Eq.~(\ref{eq:Vijn}).

\subsection{Image Synthesis with 3D baselines}
\label{sec:3Ddistribution}

\begin{figure}[htbp]
\centering
\includegraphics[width=0.4\textwidth]{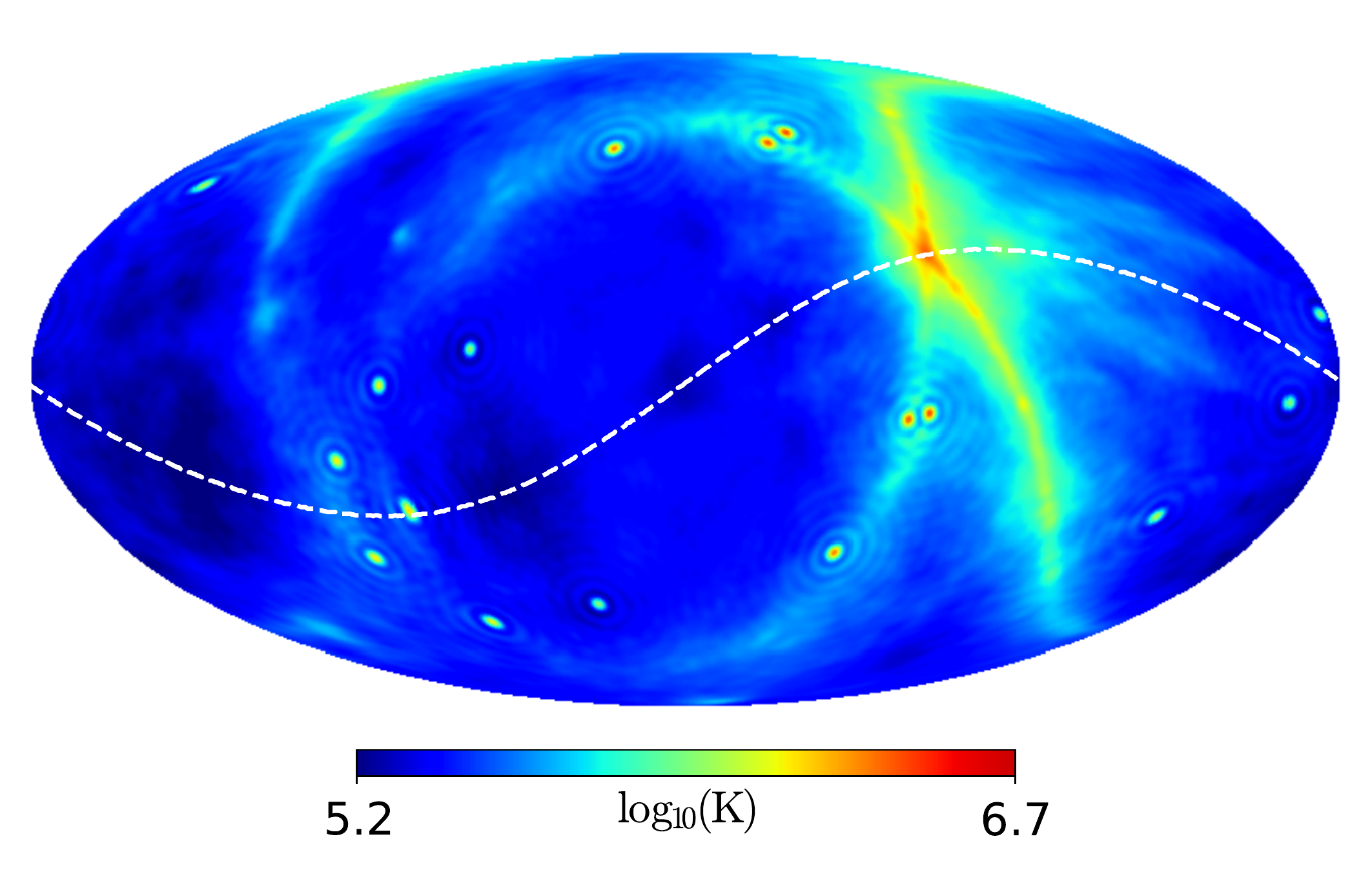}\\
\includegraphics[width=0.4\textwidth]{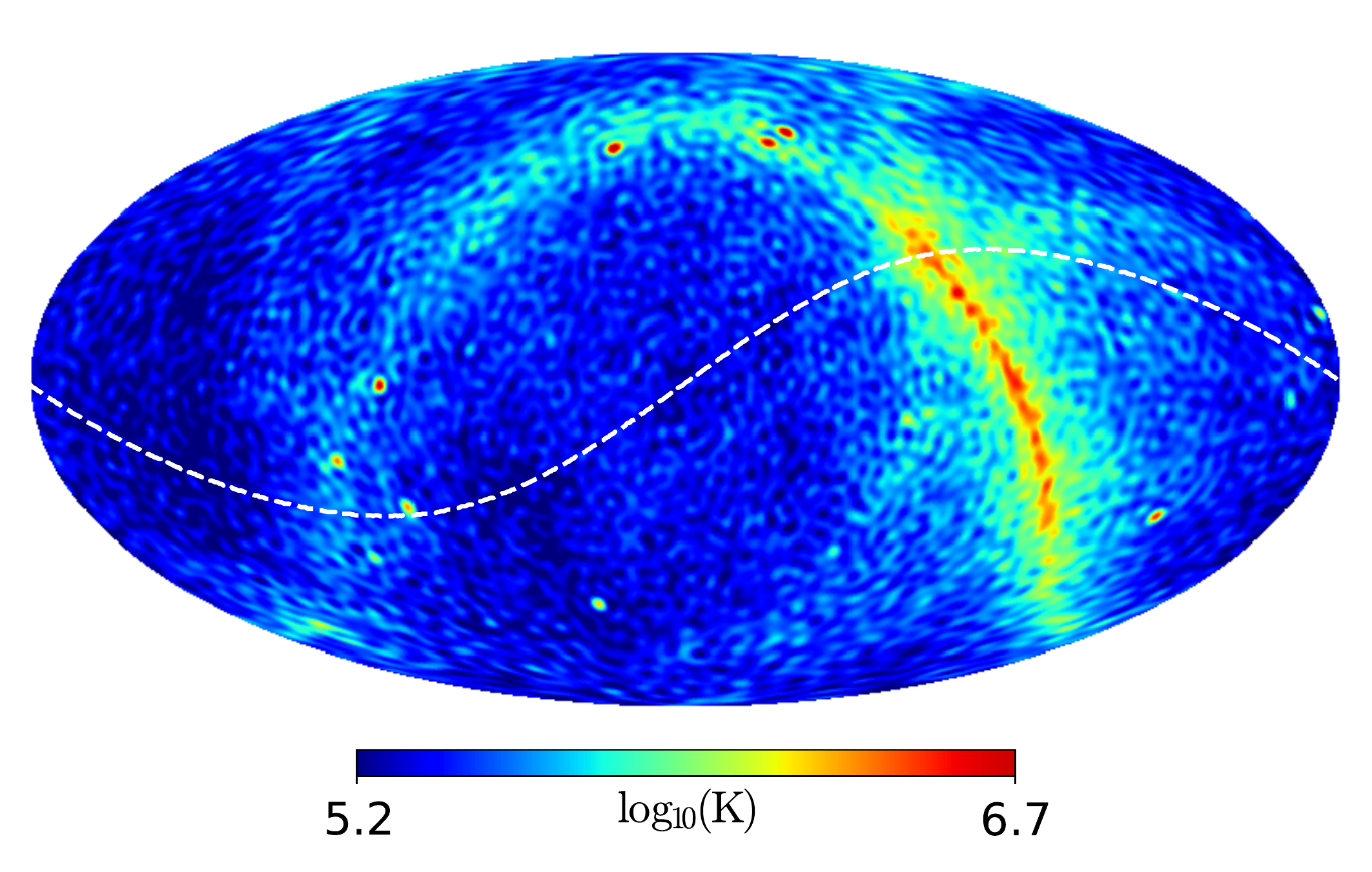}\\
\includegraphics[width=0.4\textwidth]{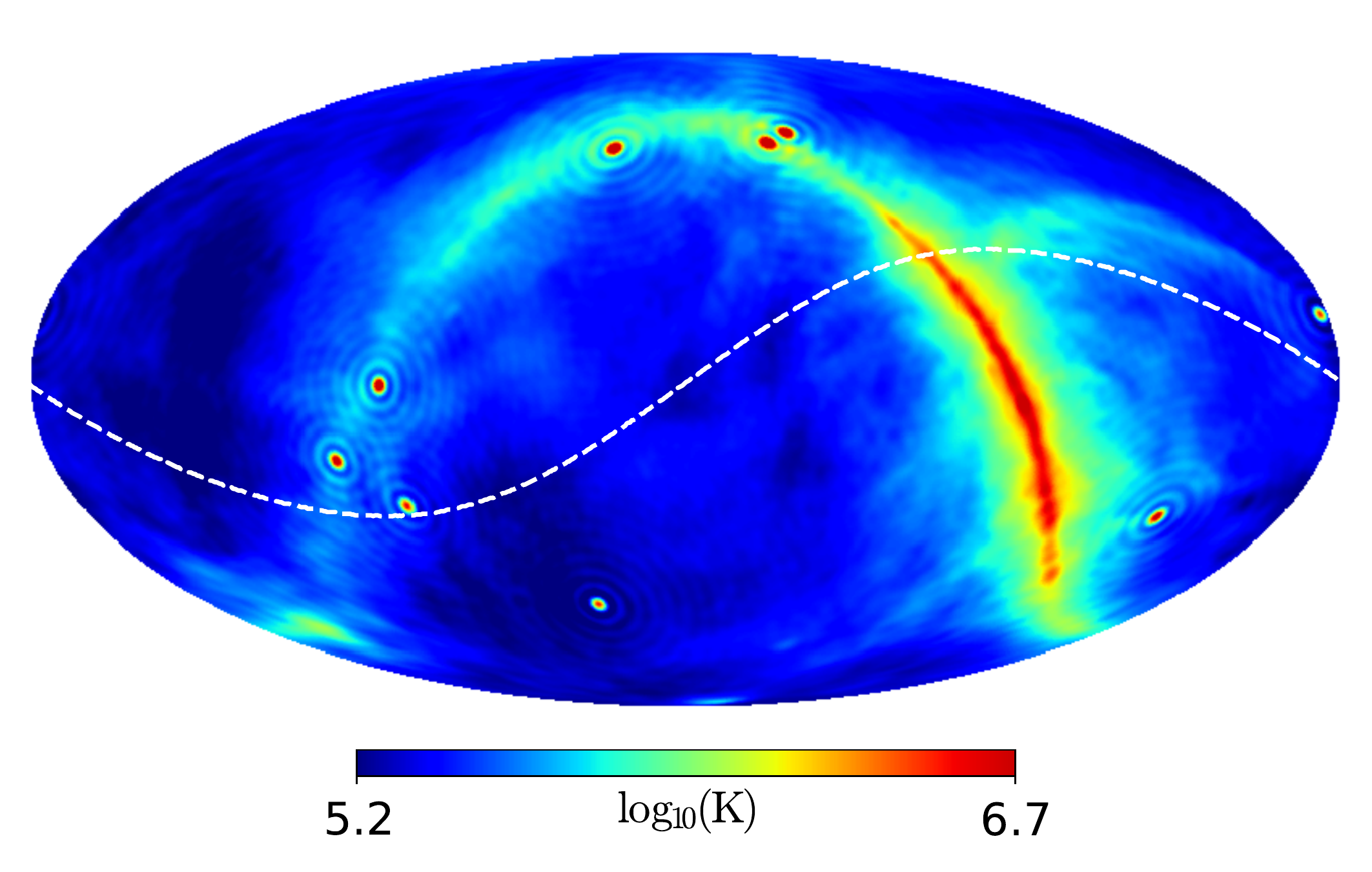}
\caption{The effect of 3D  baselines on image synthesis for the diffuse input map. 
Top: map made with all baselines on a single plane; Middle: map made with 70\% baselines on a plane and 30\% 
outside the plane; Bottom: map made with Full 3D distribution of baselines.
}
\label{fig:process_more_sky}
\end{figure}

In Fig.~\ref{fig:process_more_sky} we show the reconstruction for the input map with different 
amount of non-coplanar baselines. In each panel of this plot, the total number of visibility 
measurements are the same, but from top to bottom, the distribution is increasingly
three dimensional. The top  panel shows the image reconstructed from visibilities measured on a single plane, 
corresponding to the case of no precession. The orbit plane is shown as the dashed line in the figure. 
Here, it is quite obvious that a mirror image is formed with respect to the orbital plane, and the 
two sides are equally bright.  This is because the phase delay from either the source or its 
mirror image position are exactly the same, 
so the solution of the linear map automatically produces both.

As we mentioned earlier, this problem of mirror image can be solved by including non-coplanar baselines generated 
by the precession of the plane,  which break the symmetry.  Here, to better illustrate this effect, we temporarily 
ignore the blocking of the Moon, i.e. the Moon is treated as if it is transparent and the 
array could receive radiation from all sky directions.  In the middle panel case, the points on the single plane are 
reduced to 70\% of the total, while 30\% points are located outside the plane. We can see the mirror image is suppressed, 
though images of the brightest spots are still visible, and the quality of the map is not good due to the mixing 
of the mirror images of the diffuse structures. However, the overall structure of the true sky can already be seen. 

In the bottom panel, we show the case of the fully 3D view, i.e.  the visibilities are measured in the full 3D space,
here the mirror images disappeared, and the 
reconstructed map is very close to that of the original, showing that with the  baselines distributed 
over three dimensions good maps could be made.

\subsection{The Effect of Blockage by the Moon}

As the satellites fly around the Moon, at any moment the Moon blocks part of the sky. The direction of the blocking is 
position-dependent: the central direction is toward the center of the Moon, which is perpendicular to the tangent 
of the orbit, and for a linear array on the circular orbit this also nearly coincides with the direction of the 
baselines between the satellites. As a result, for each baseline, the visibility is obtained for a sky which is 
screened slightly differently. However, since we know the exact position of the region being blocked, 
the sky map can still be recovered with the brute-force algorithm described in Sec. 2.

\begin{figure*}[htbp]
\centering
\includegraphics[width=0.4\textwidth]{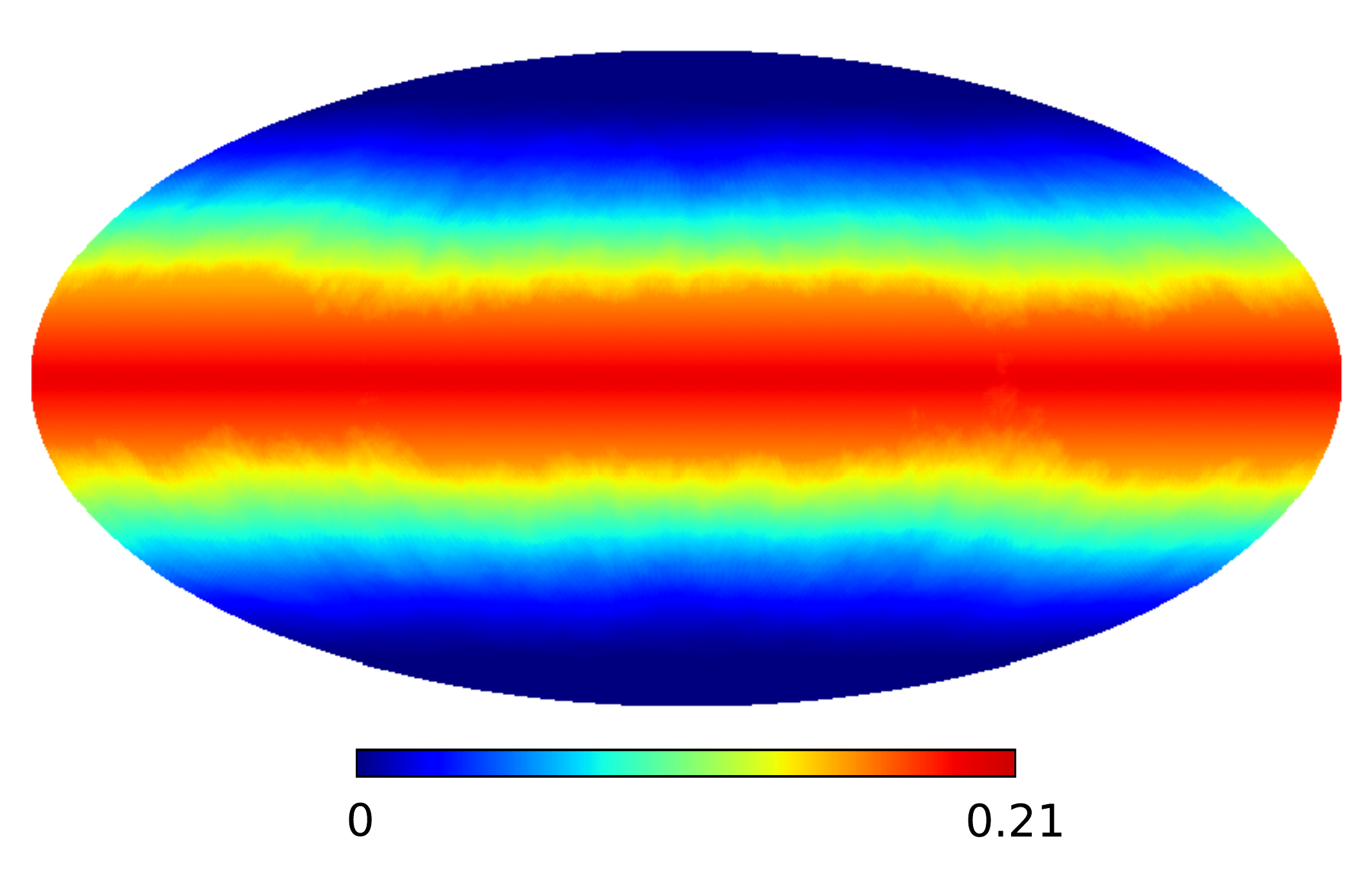}
\includegraphics[width=0.4\textwidth]{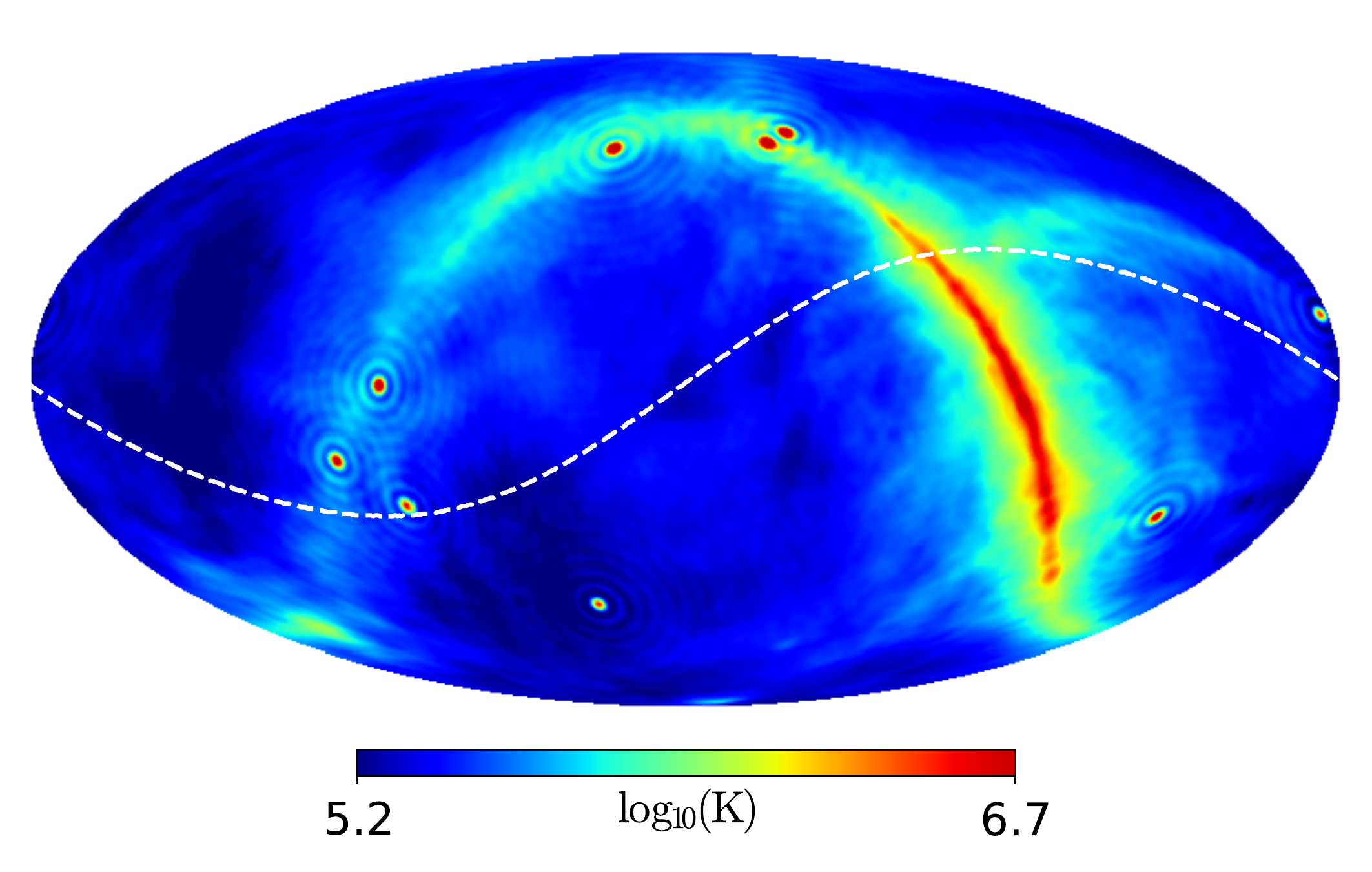}
\caption{The effect of Moon blockage. Left Panel: the blockage fraction for each direction on the sky; 
 Right Panel: the corresponding reconstructed image for the diffuse map. }
\label{fig:blocking_source}
\end{figure*}

In Fig.~\ref{fig:blocking_source}, we investigate how the effective blockage fraction (left panel) 
affect the reconstructed  map (right panel). The effective blockage fraction is 
the fraction of measurements  being blocked by the Moon in that direction.  That is, for each pixel in the map, 
we count what is the fraction of the the mock visibility sample data taken when this pixel is blocked by the Moon. 
As a result of orbit plane precession, the blocking of the Moon moves across the celestial sphere, so only in 
a fraction of measurements that part of the sky is blocked. The blockage fraction is nearly uniform along
the ecliptic longitude, with about 21\% blocked at the maximum near the ecliptic plane, and drops to zero
near the ecliptic pole. The baselines are generated randomly, so there is some slight fluctuations in the blocking fraction.
As shown in the right panel, the map is well reconstructed and unbiased despite the non-uniformity in the latitudes.

\subsection{Pixels vs. Spherical Harmonics}

In Sec. 2, within the general framework of image synthesis by solving linear mapping equation, 
we introduced two different methods. One approach is to solve the equations linking the visibilities directly to the 
intensity distribution on the sky pixels (Sec.\ref{sec:pixel}),  the other is to replace the pixels by spherical harmonic 
coefficients (Sec.\ref{sec:ourmethod}). Here we compare the reconstructed maps from these two methods. 
As the FWHM resolution of our input map is $1^{\circ}$, we choose a lower resolution for  the output map, so that 
assessment of the quality of the output map would not be affected by the resolution of the input map.
For the pixel method, we choose HEALPix $n_{\rm side}=32$, with pixel size $1.83^{\circ}$. 
For the spherical harmonic method, we choose $l_{\rm max}=98$, whose corresponding resolution 
is the same as $n_{\rm side}=32$. 

\begin{figure*}[htbp]
\centering
\includegraphics[width=0.4\textwidth]{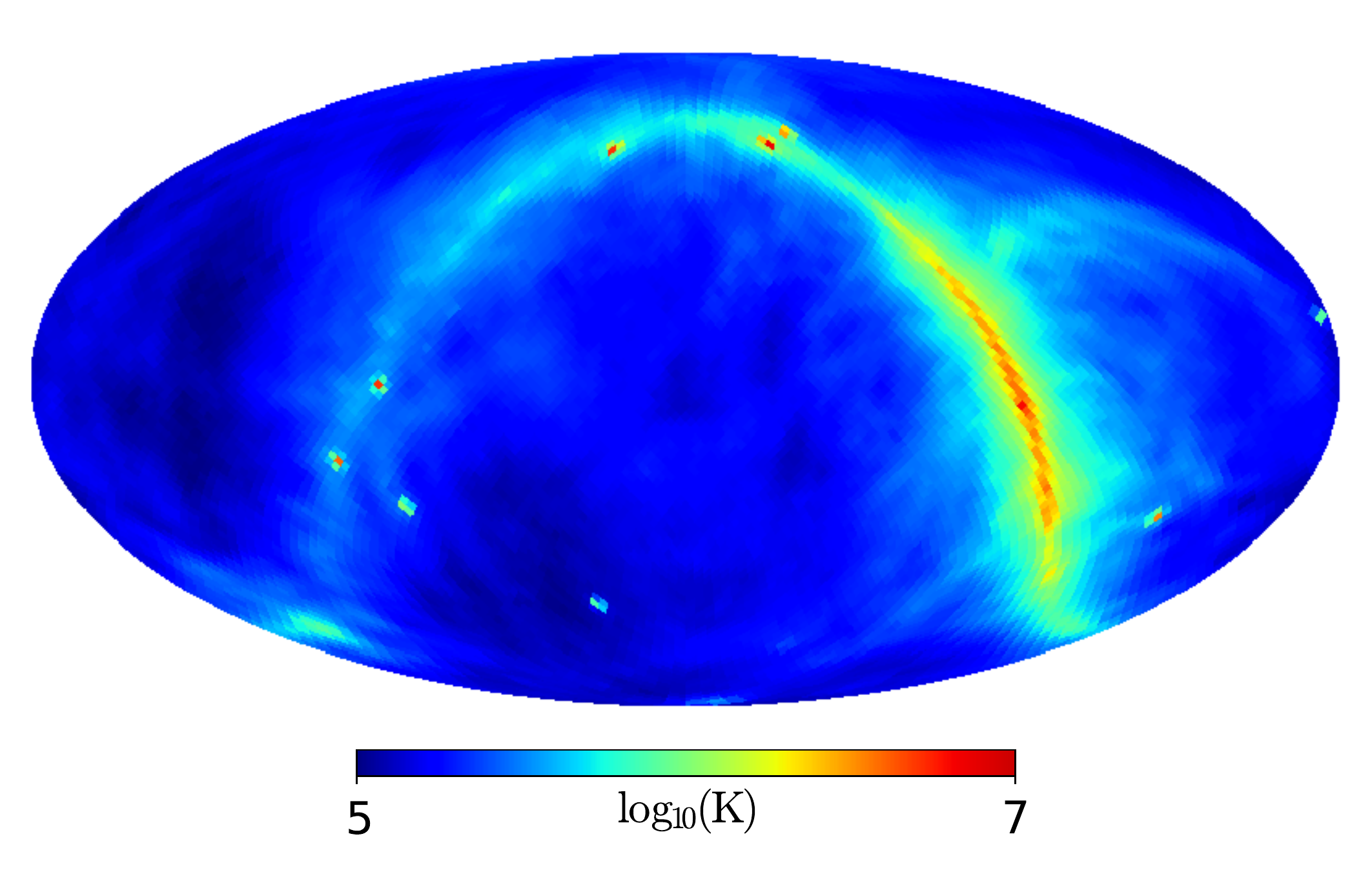}
\includegraphics[width=0.4\textwidth]{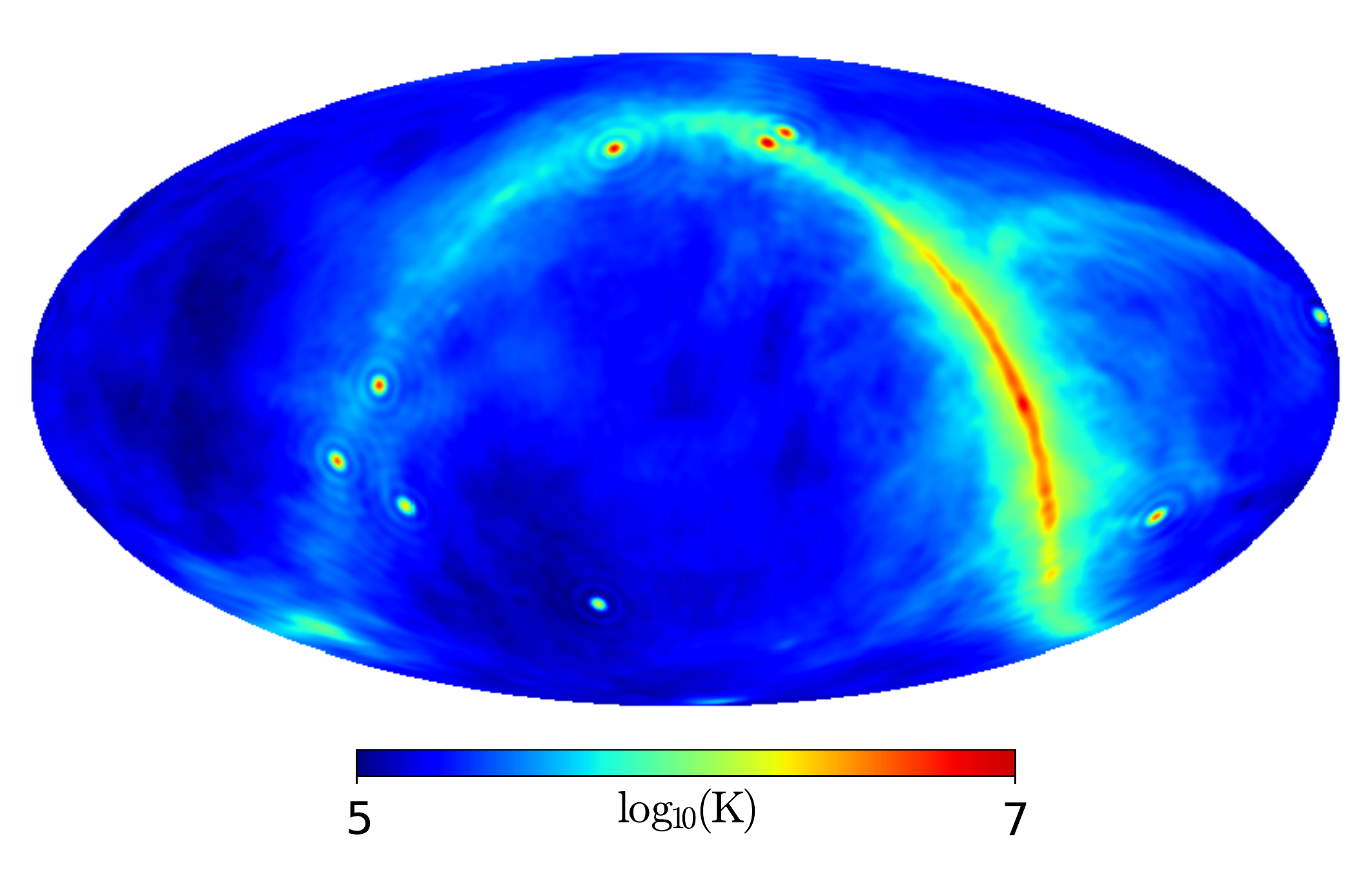}\\
\includegraphics[width=0.4\textwidth]{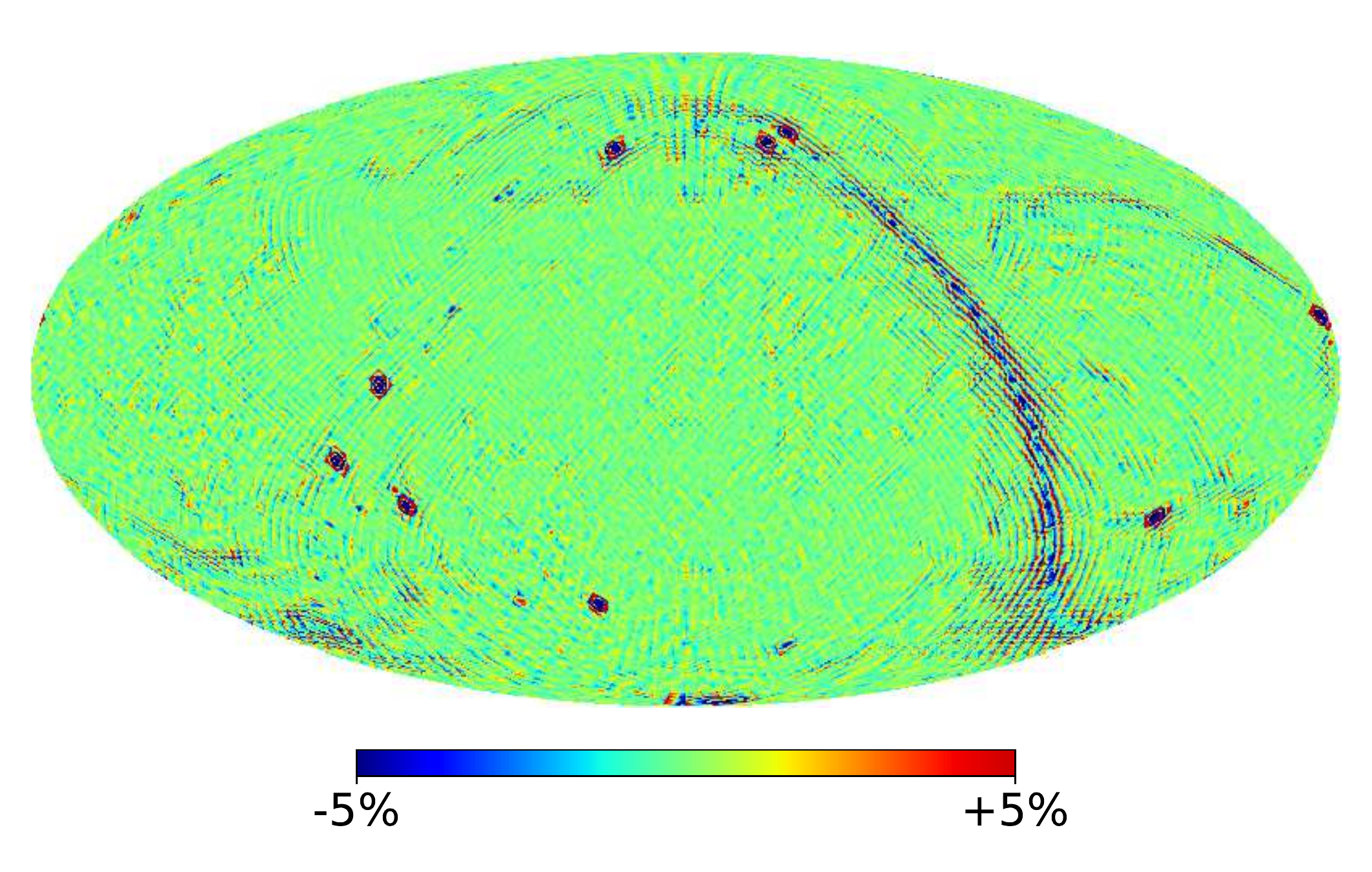}
\includegraphics[width=0.4\textwidth]{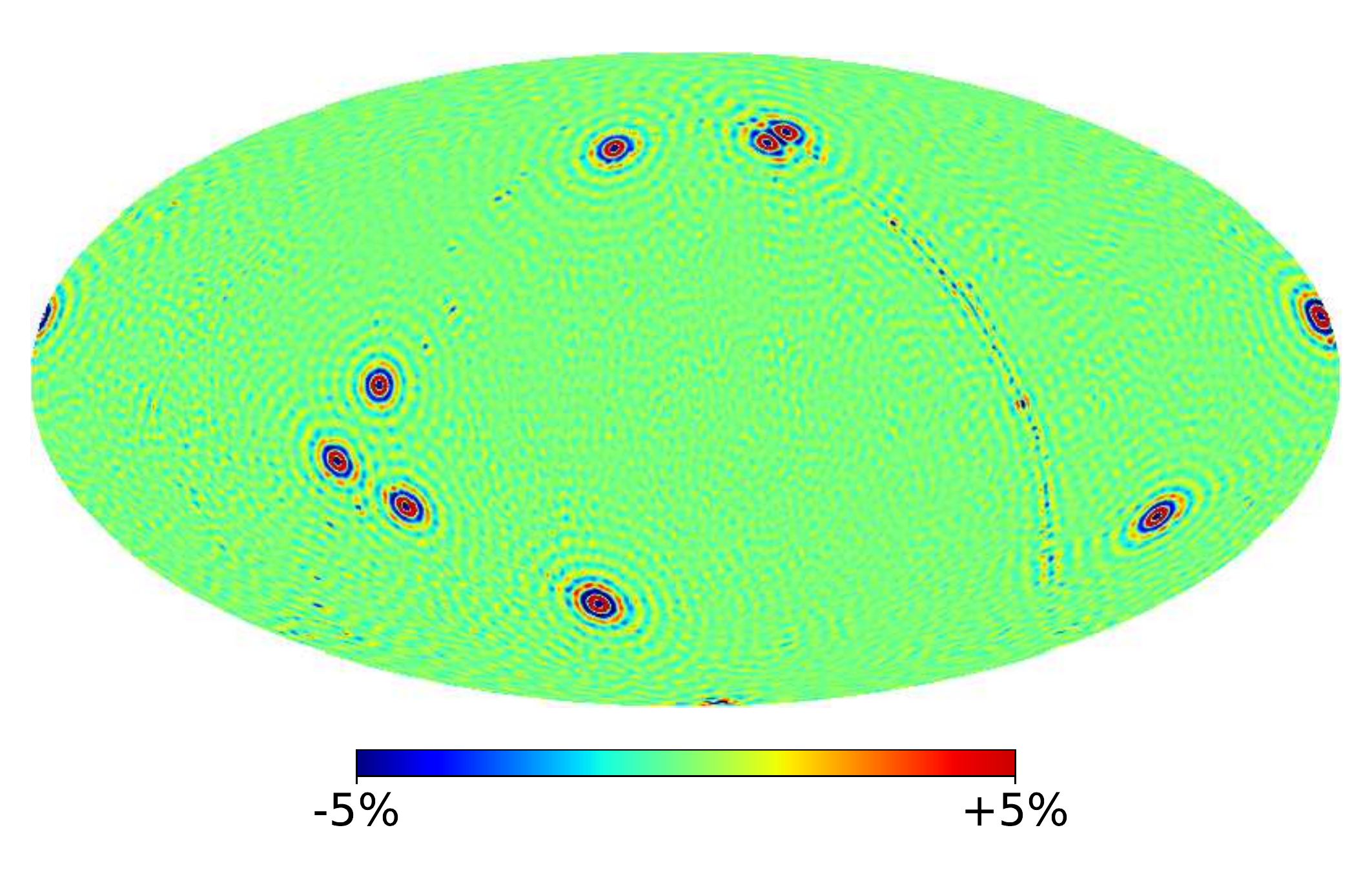}
\caption{The reconstructed maps (top) and their relative errors(bottom). Left: map reconstructed in the pixel base.
Right: map reconstructed in the spherical harmonics base. For the relative error map, 
 we limit the colorbar from -5\% to +5\%, but the maximum, minimum, mean and median errors are 
 $\sim$45\%, $\sim$0.02\%, $\sim$2.3\% and $\sim$1.7\% respectively for the pixel base method, and 
 $\sim$20\%, $\sim$0.006\%, $\sim$1.6\% and $\sim$1.2\% respectively for the spherical harmonic method.}
\label{fig:maperror}
\end{figure*}

In Fig. \ref{fig:maperror} we show the reconstructed maps (top panels) and the relative errors (bottom panels) 
for the two methods.  In both cases, the relative error are largest at the brightest sources. Due to the imperfect 
reconstruction,  the much higher brightness temperature at these point are leaked into neighboring pixels, 
so the peak of a source are lower than its true value, while its neighboring pixels are raised to higher temperature
than their true values. For similar reason, the relative error on the Galactic plane is also larger. 

When comparing the two methods, we see the relative errors of the map made with the pixel base method 
are obviously much larger. The maximum absolute value of the relative error is $\sim$45\%,  
the mean error is $\sim$2.3\% and the median error is $\sim$1.7\%. 
The map made with the spherical harmonics method has smaller errors, with the 
maximum, mean and median errors at  $\sim$20\%, $\sim$1.6\% and $\sim$1.2\% respectively.

\begin{figure}[htbp]
\centering
\includegraphics[width=0.4\textwidth]{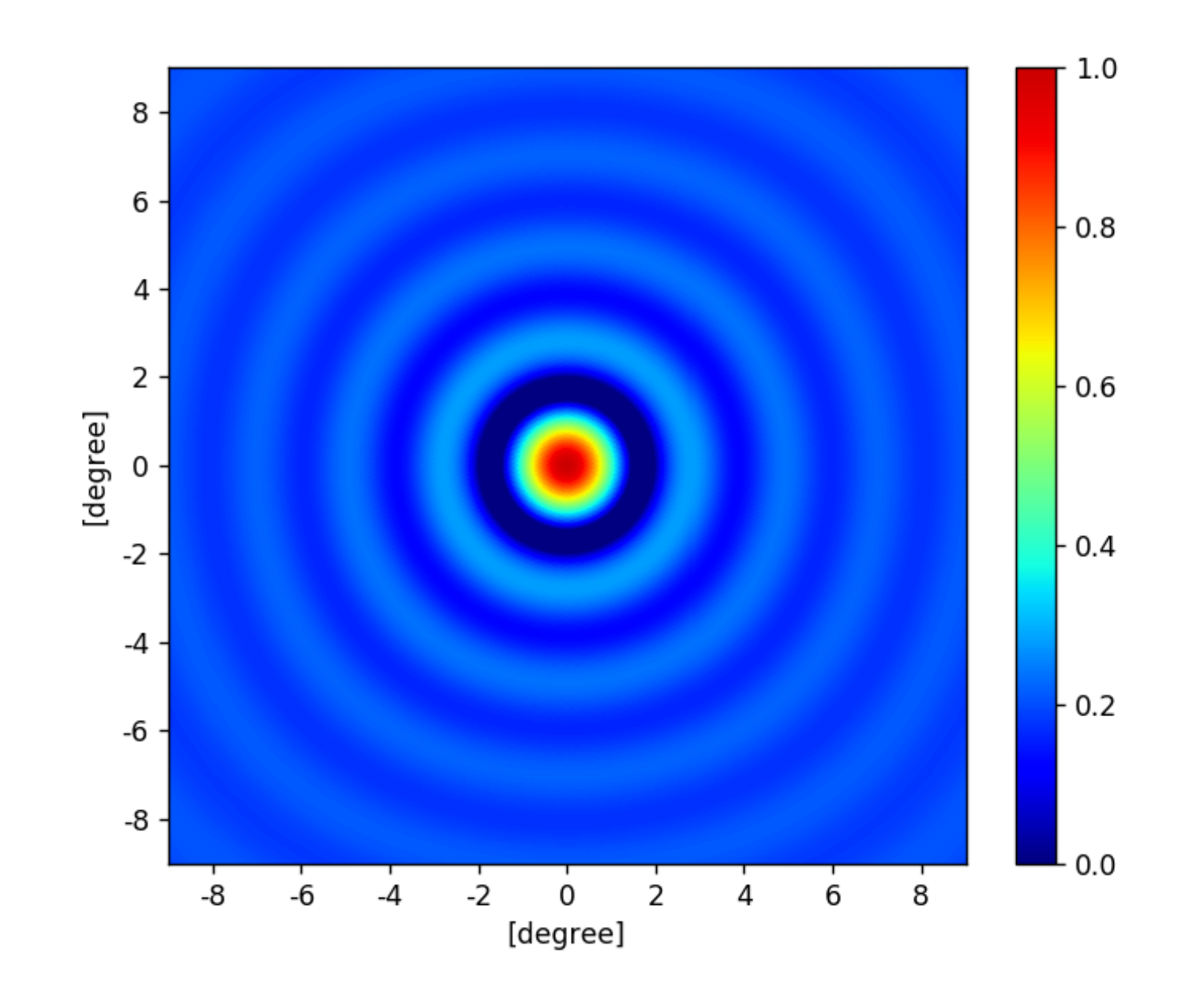}
\includegraphics[width=0.4\textwidth]{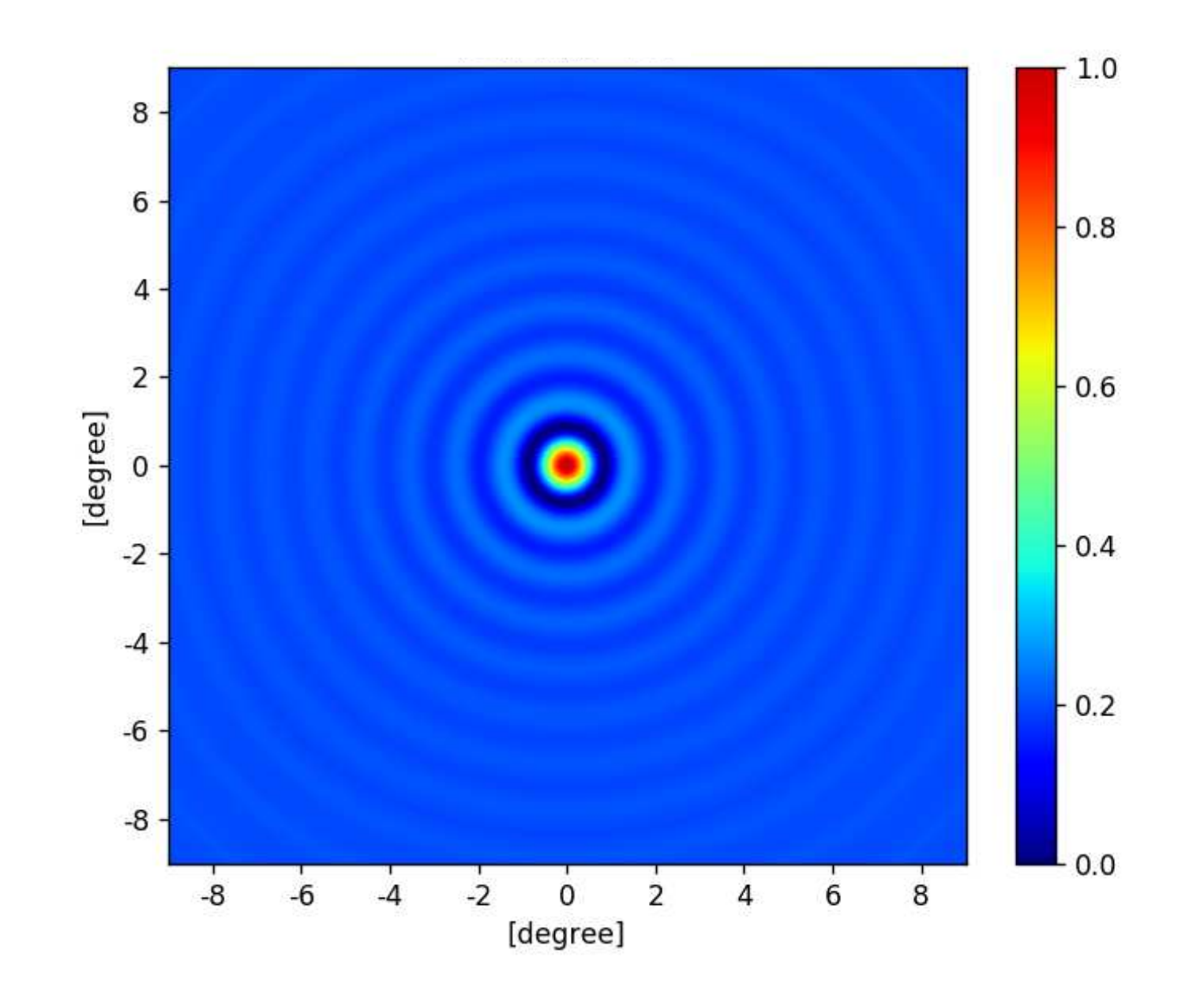}
\caption{The effect of $l_{\rm max}$ on PSF. Top: $l_{\rm max}=100$;  Bottom: $l_{\rm max}=200$. }
\label{fig:psf}
\end{figure}

When expanding the sky intensity to the spherical harmonics, because in practice the expansion is always finite
and limited to $l_{\rm max}$, some information on the small scales are lost. For point sources, the peak intensity is lower, 
and we could see rings of side lobe appears around the points.  The point spread function for $l_{\rm max} =100 $ and 200
are plotted in Fig.~\ref{fig:psf}. For the higher $l_{\rm max}$, the size of the central peak and the side lobes are all smaller. 
However, fundamentally the PSF is determined by the baseline distributions.  
The side lobes could be reduced if we replace the sudden cut off in $l$-mode by a soft cut off, 
e.g. if an exponential damping function is applied at $l$. 

\subsection{The effect of baseline distribution}

\begin{figure}[tbp]
\centering
\includegraphics[width=0.45\textwidth]{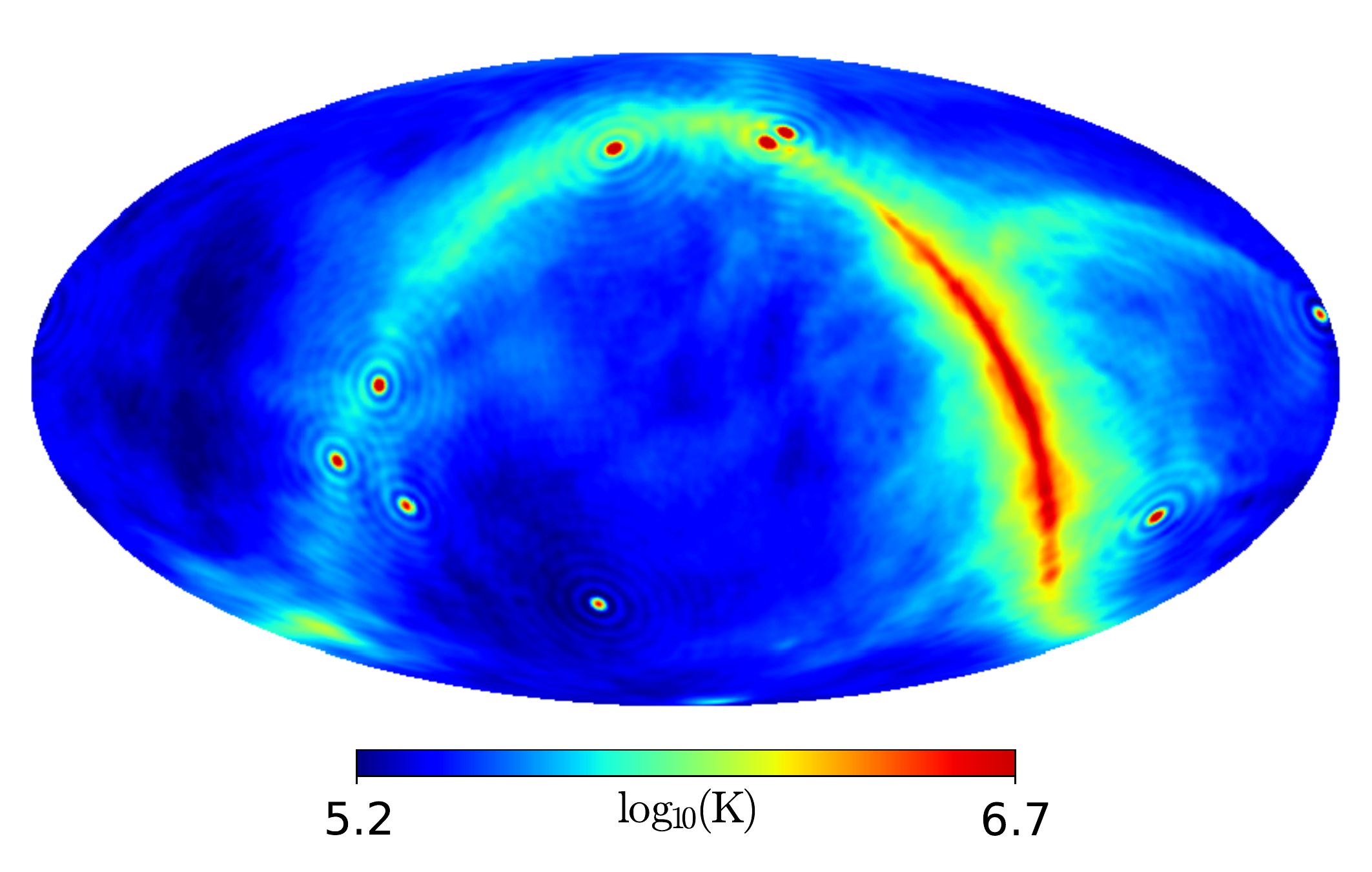}
\caption{The synthesized map with only relatively long ($6\km < |\vec{u}| < 10 \km$) baselines.}
\label{fig:longbaseline}
\end{figure}

In the lunar orbit array mission, each baseline makes a circle or ellipse track during one orbit. Sampling 
of the $(u,v,w)$ space is achieved by placing the array satellites at different distances, and by adjusting 
their distances during the operation. Since there is only a finite number of satellites and also
limited time for operation, a complete sampling of the $(u,v,w)$ space is impossible. However, 
in radio astronomy, thanks to the nature of sky radiation (i.e. most of the sky radiation power are concentrated in  
point sources or a limited number of features), sky image could still be made with observations which only sparsely
sample the $(u,v,w)$ space.  In the simulations shown above, the sampling on the $(u,v,w)$ space is 
far from complete: at our fiducial 10 MHz frequency, $u_{\rm max}$$\sim$300, and 
for about $8 \times 10^3$ sampling points, the sampling density is about $10^{-3}$, yet the reconstructed map 
has reasonably good quality. Furthermore, in Fig.~\ref{fig:longbaseline} we show the map made by using only  
the relatively longer baselines  ($6 \km \sim 10\km$), i.e. there is a "hole" in the center of the 
3D baseline distribution. Due to the missing short baselines, the sampling of the large angular scale structures are 
incomplete. Nevertheless, visually the large structures such as the galactic plane are still reproduced
with reasonably good quality. The side lobes however become stronger in this case due to the missing short baselines.

For a lunar orbit array, the visibility data would be acquired over an extended period of time.
In the initial stage of its operation, due to limited amount of precession, 
the sampling over the $(u,v,w)$ space is over a thin disc, similar to the cases we simulated in 
Sec.~\ref{sec:3Ddistribution}.  With more time, the sampled points in the $(u,v,w)$ space would 
round up, however the distribution may still be inhomogeneous due to a variety of practical reasons.

\begin{figure}[htbp]
\centering
\includegraphics[width=0.4\textwidth]{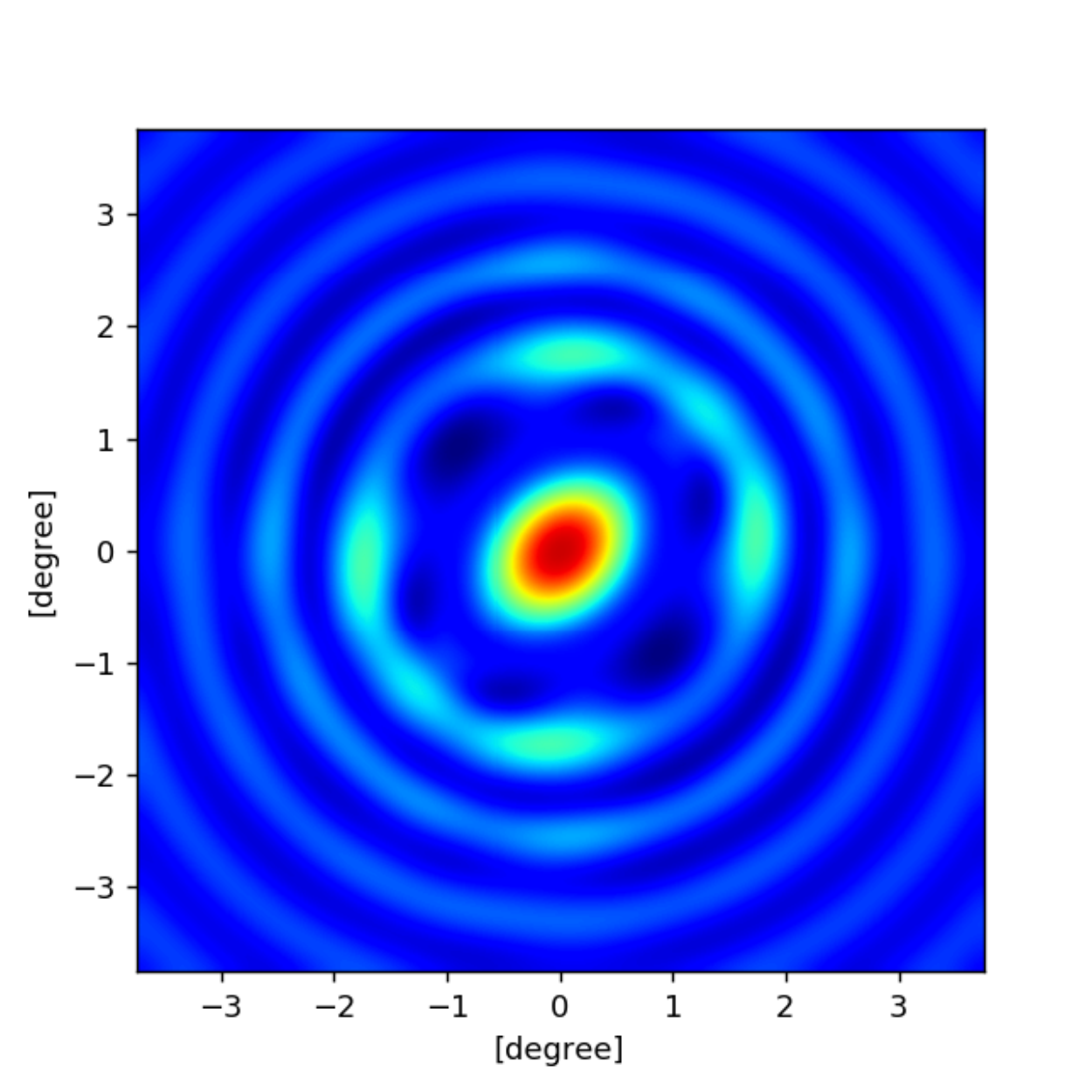}
\includegraphics[width=0.4\textwidth]{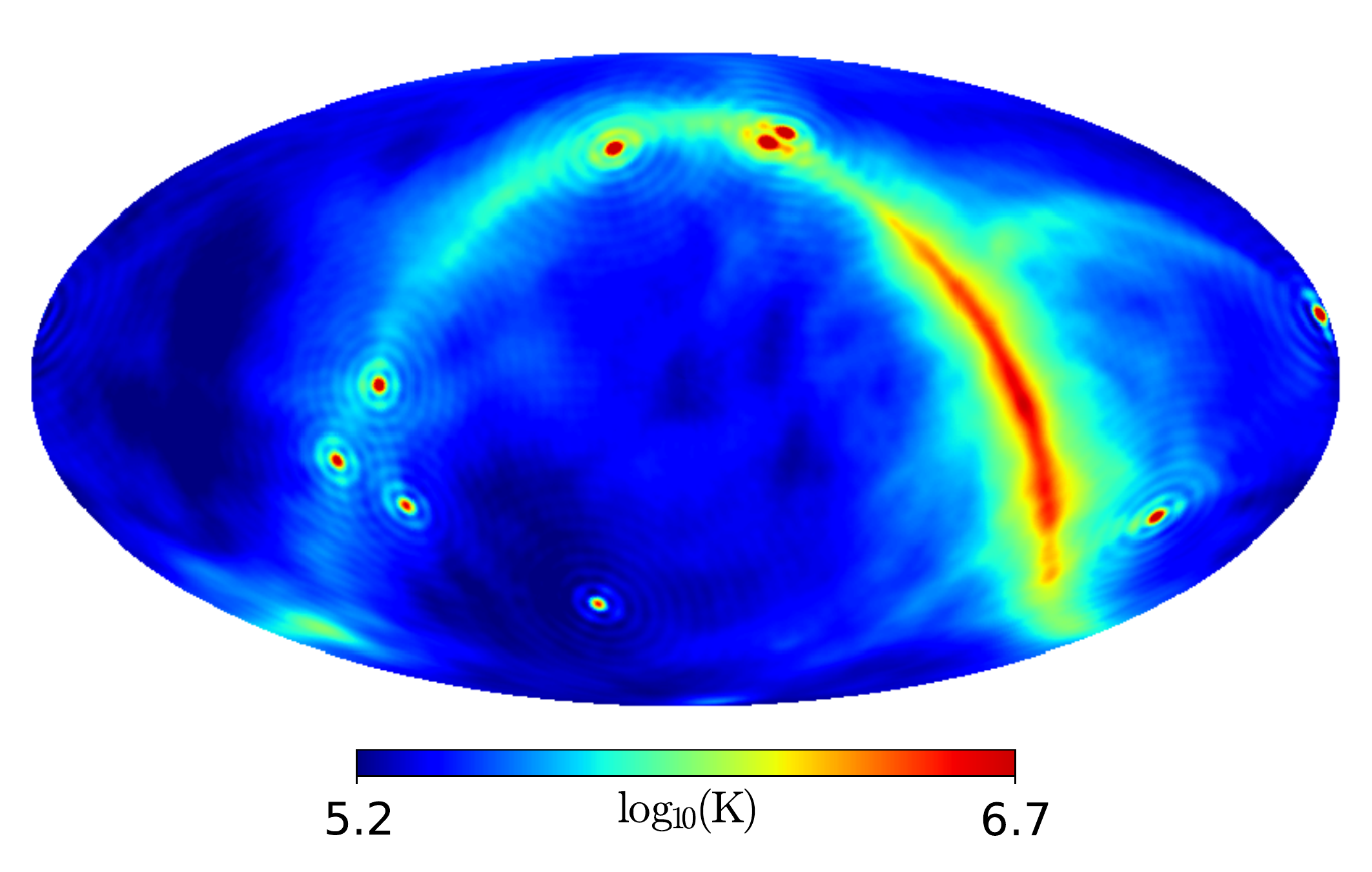}
\caption{Image synthesized with asymmetrically distributed measurements, where all visibilities are taken only in the
longitudinal range of $0^\circ$--$90^\circ$. 
Top: A single point source zoomed up;  Bottom: the constructed map for the whole sky.}
\label{fig:baseline_number_source}
\end{figure}

In Fig.~\ref{fig:baseline_number_source} we show the impact of the inhomogeneous distribution in an extreme case, 
where we limit the baselines to be acquired only in the longitudinal range $0^\circ$--$90^\circ$, i.e. 1/4 of the whole
orbital distribution shown before. Note however that if the operation time is limited and one wants to use only the data 
taken in the dark side of the Moon (i.e. when both the Sun and Earth are blocked) this may well be the case.
The top panel shows a zoom in on a point source, where we can 
see that the shape of the central image is a bit distorted, and there is some asymmetric variations of 
brightness in the side lobe rings.  The bottom panel shows the impact on the whole diffuse structure of 
the map. There is some distortion in the reconstructed points, but even in this extreme 
case we could still obtain a reasonably good map of the sky. 

\section{Discussions and Conclusion}
In this paper we investigate the image synthesis for a lunar orbit interferometer array. Such an array 
is ideal for observing the sky at frequencies below 30 MHz, which is difficult to do on the ground due to 
the ionosphere effects and RFIs. Compared with an array on the far side of the lunar surface, the orbiting array
does not require landing on the Moon and deploying the array elements at different locations, 
there is no need for a separate relay satellite to transmit the data back to Earth, and the power can be 
readily provided by conventional solar battery involving standard technology in space operations. 

However, the orbiting array does require interferometry and image synthesis techniques different from the ground-based 
arrays, because (1) due to practical considerations, at this frequency band electrically short antenna is almost the only choice, 
so the satellites would have large (almost whole sky) field of view; (2) only a small fraction of the sky is blocked
by the Moon at any time, and there is a mirror symmetry for the image synthesized from visibilities measured on a single
plane. Because of these differences, the usual small-$w$ approximation used in the synthesis imaging of ground arrays 
which requires small FoV and planar array is not applicable, and indeed three dimensional distribution of baselines
are desirable to break the mirror symmetry.  Furthermore, in the orbital linear array, the baseline direction and the 
position of the array relative to the Moon are correlated, so for different baselines, the blocked sky region are also 
different, none of the traditional imaging algorithms is applicable in this case.

In this paper we propose to  make map of the sky by numerically solving the measurement equations which relate 
the visibility to sky intensity,  the baseline-dependent screening is then automatically dealt with.
We also uses spherical harmonic decomposition of the sky intensity, which can speed up 
the computation in some cases and and produces higher quality maps.
Our simulations show that the sky map could be recovered very well using this method. 

The present study is only a first step in this direction, which demonstrated the general idea, but we have made many 
simplifications and ignored a lot of practical issues. We only considered the interferometry and completely ignored the 
antenna beam pattern and polarization. Generalization to include these are straightforward, 
which we plan to do in the next step. 
Another simplification is that we assumed the same patch of sky is being screened by the Moon for all satellites, but 
in fact due to their different positions there is some slight difference for each one.
We also neglected the reflection and diffraction of the Moon, and treated the Moon as a simple sphere. In fact, the 
radio waves would be reflected by the Moon, and there are rugged terrains on the Moon, which will all affect the 
signals received by the array. We assumed a static sky during the time of observation, but the Sun and planets such as 
Jupiter do have variations, which may affect the image reconstruction. 

Another omission in this paper is calibration, which is of fundamental importance to the working of 
interferometer array. For the orbiting array, there are inevitably variations in the length of baselines 
and relative velocities at all times, which also affect the synchronization of
time and frequency reference on the satellites. Also, sky-based calibration may have special problems at this 
wavelength, because usually the first step of calibration is to use a single bright source as calibrator,  though more 
sophisticated methods do exist. 
However, in the present case the field of view of the antenna is nearly the whole sky, there would be many sources present 
in the field of view at any time, and the sky may also be relatively bright at low frequencies. 

These problems are all very important, but they are beyond of the scope of the present work. 
We shall  make further investigations on  these problems in subsequent studies.

%%%%%%%%%%%%%%%%%%%%%%%%%%%%%%%%%%%%%%%%%%%%%%%%%%
\acknowledgements

We thank our colleagues from the DSL team and the CE-4 piggybacking long wavelength astronomy 
micro-satellites team for many helpful discussions during the past several years which enhanced our 
understanding of the various aspects of the lunar orbit long wavelength array. 
This research is supported by the CAS Strategic Priority Research Program XDA15020200, the CAS 
Frontier Science Key Project QYZDJ-SSW-SLH017, the NSFC key project grant 11633004 and grants 11473044,
11653003, 20171352322 and U1501501, and the MoST 2016YFE0100300. 
F. Q. Wu also benefitted from the support of the CSC Sino-French Cai Yuanpei travel grant.

%%%%%%%%%%%%%%%%%%%%%%%%%%%%%%%%%%%%%%%%%%%%%%%%%%
\bibliographystyle{hapj}

\bibliography{moonsatellite}

\end{document}